\documentclass[conference]{IEEEtran}
\IEEEoverridecommandlockouts
\usepackage{cite}
\usepackage{amsmath,amssymb,amsfonts,mathtools}
\usepackage{algorithmic}
\usepackage{booktabs}
\usepackage{subcaption}
\usepackage{graphicx}
\usepackage{textcomp}
\usepackage{xcolor}
\usepackage{url}
\usepackage{hyperref}
\def\BibTeX{{\rm B\kern-.05em{\sc i\kern-.025em b}\kern-.08em
    T\kern-.1667em\lower.7ex\hbox{E}\kern-.125emX}}

\hypersetup{
    pdfauthor={Josue N Rivera},
    pdftitle={Scalable Online UAM Trajectory Optimization via SQP in Ultra Low-Altitude Airspace},
    pdfcreator={Josue N Rivera}
}
\begin{document}

\title{Scalable Online Flight Trajectory Optimization via Sequential Quadratic Programming for Urban Air Mobility in Ultra Low-Altitude Airspace\\
\thanks{This research was supported by A*STAR under its RIE2025 Manufacturing Trade and Connectivity (MTC) Industry Alignment Fund - Pre- Positioning (IAF-PP), with Award No. M23L5a0002.}
}

\author{\IEEEauthorblockN{Josue N. Rivera, Bohang Liang, Chen Lv, and James Wang}
\IEEEauthorblockA{\textit{School of Mechanical and Aerospace Engineering} \\ 
\textit{Nanyang Technological University}\\
Singapore, Singapore \\
\{josue.rivera, lyuchen, james.wang\}@ntu.edu.sg, bohang001@e.ntu.edu.sg}
}

\maketitle

\begin{abstract}
As Urban Air Mobility (UAM) scales toward high-density operations, generating collision-free trajectories within complex 3D cityscapes is a critical safety requirement. This paper proposes a scalable Sequential Quadratic Programming (SQP) framework that integrates geometric environmental constraints, operational limits, and vehicle dynamics within a single online trajectory optimization process. Rather than precomputing obstacle-free corridors ahead of time, our method encodes obstacle avoidance as live separating-hyperplane constraints regenerated at every solver iteration, so that dense urban geometry and full-DOF vehicle dynamics are resolved jointly and online as the reference and environment evolve. A variable-scale quadtree decomposition keeps computation bounded, enabling the framework to scale to city-wide environments while preserving real-time performance for high-speed flight. We validate the framework against conventional SQP, Iterative Linear Quadratic Regulator, and Differential Dynamic Programming across flights in five real-world urban centers, attaining 100\% success and clearance rates on CPU-only hardware.
\end{abstract}


\section{Introduction}\label{sec:intro}

The promise of Urban Air Mobility (UAM) lies in reimagining city life through the third dimension, offering efficient and reliable movement of people, goods, and services across congested cities~\cite{li2024urban}. In compact, high-density environments such as San Francisco, New York, Singapore, or Hong Kong, UAM offers a compelling solution to ground-level saturation that conventional transportation cannot resolve. However, broad deployment remains impeded by concerns over community noise, privacy, and, most critically, operational safety~\cite{al2020factors, charnsethikul2025urban}.

Safety concerns are particularly acute in ultra low-altitude airspace (below 150\,m above ground level, AGL), where vehicles share confined urban volumes with buildings and other aircraft. At these altitudes, the risks of midair collision, infrastructure strike, and third-party ground impact are substantially elevated relative to conventional aviation~\cite{li2026mid}. For aerial applications to scale, trajectory generation must therefore simultaneously guarantee collision avoidance with real-world infrastructure and dynamic feasibility with respect to vehicle limitations. Yet many real-time implementations either treat geometric collision avoidance as a preprocessing step separate from the dynamic optimization (which can lead to suboptimal solutions) or only encode a small, fixed set of obstacles (which can lead to safety-margin violations in dense environments and limited scalability as the number of obstacles grows)\cite{liu2017planning, singh2022Optimizinga, howell2019altro}.

This paper proposes the Last-Mile Trajectory Planning (LTP) framework to address the challenge of generating safe and dynamically feasible trajectories in dense urban settings. LTP formulates the problem as a Sequential Quadratic Program (SQP) that jointly optimizes over vehicle dynamics, operational constraints, and real-world geometry. By encoding obstacle avoidance as an intrinsic live constraint, the planner autonomously corrects for ill-conditioned global references and challenging demands. Furthermore, a variable-scale cell decomposition keeps computation bounded, enabling the framework to scale to city-wide environments while maintaining real-time performance for high-speed flight.

The contributions of this paper are fourfold:
\begin{itemize}
    \item \textbf{Live In-the-Loop Geometric Avoidance:} We formulate an online SQP trajectory optimizer that encodes building avoidance as live separating-hyperplane constraints regenerated at every solver iteration, rather than as precomputed corridors, so that dense urban geometry and full-DOF vehicle dynamics are resolved jointly within a few Quadratic Programming (QP) iterations.
    \item \textbf{Scalable Spatial Decomposition:} We introduce a variable-scale quadtree decomposition over an Unmanned Aircraft System Traffic Management (UTM) channel. This restricts each inner QP call to a locally relevant constraint subset along a trajectory, maintaining near-constant solve times even as the environment scales to city-wide maps with hundreds of obstacles.
    \item \textbf{High-Performance Digital Twin Pipeline:} We develop a JAX-accelerated digital-twin model that fuses automatic differentiation, zero-order-hold discretization, and JIT compilation into a single XLA kernel. This enables sub-millisecond discretization, and supports $10$--$20$\,Hz replanning on CPU-only edge hardware.
    \item \textbf{UAM Trajectory Planning Benchmark and Open-Source Release:} We validate the framework across a comprehensive benchmark of $1710$ flights (across the various algorithms) in five real-world urban environments, and release the source code to support UAM research reproducibility.
\end{itemize}

The remainder of this paper is organized as follows. Section~\ref{sec:related} situates the framework within the urban air mobility, unmanned aircraft system traffic management, and trajectory-optimization literature. Section~\ref{sec:ltp} formalizes the LTP framework, covering the digital-twin dynamics model, the live half-plane geometric constraints, the variable-scale quadtree decomposition, and the QP transformation to standard form for SQP\@. Section~\ref{sec:results} reports the simulations benchmark, comparisons against DDP, iLQR, and SQP baselines, and two city-scale case studies in Singapore and Boston. Section~\ref{sec:conclusion} concludes with the insights found and discusses extensions toward a full UAM stack.

\section{Related Works}\label{sec:related}

\subsection{Urban Air Mobility and Traffic Management}
Urban air mobility has emerged over the last decade as a proposed solution to ground-level congestion in dense metropolitan areas, promising efficient movement of passengers and cargo through low-altitude airspace~\cite{li2024urban, moradi2024Urban}. Its realization is contingent on resolving a tightly coupled set of economic, social, and operational challenges, from community noise and equitable network planning to conflict management and autonomous cargo certification~\cite{al2020factors, barsotti2026integrating, gao2024noise, groll2025integration}. Operational safety is among the most critical of these challenges. At low altitudes, UAM vehicles share confined urban volumes with buildings and other aircraft, creating substantially elevated risks of midair collision, infrastructure strike, and third-party ground impact compared to conventional aviation~\cite{charnsethikul2025urban, li2026mid}.

Addressing these challenges at scale requires a dedicated unmanned aircraft system traffic management (UTM) stack for strategic deconfliction, tactical separation, and trajectory supervision~\cite{pintoneto2022Trajectory, rivera2024air}. Research in UTM has yielded three primary morphologies for structuring low-altitude airspace: gridding, stratification, and networking~\cite{yang2024Review}. These approaches partition operational volumes into structures such as corridors, tubes, or lanes, each offering a different trade-off between flight flexibility and traffic predictability~\cite{yang2024Review, jang2017concepts, bauranov2021designing}. Our framework adopts the altitude \emph{stratification} model, where the airspace is segmented into distinct altitude layers within which aircraft can navigate freely. The horizontal partition is driven by factors such as buildings height, mission needs, vehicle classes, and regulations.

\subsection{UAV Navigation System}
Autonomous UAV navigation is typically realized through a modular architecture comprising four tightly coupled stages: perception (e.g., sensor fusion from LiDAR, cameras, IMU, and GPS to build environment representations), obstacle state estimation and prediction (tracking dynamic objects and forecasting future positions), motion planning (computing optimal trajectories), and low-level control (tracking the planned trajectory with actuator commands)~\cite{hashim2025advances}. This modular decomposition allows each component to be developed, tested, and certified independently, which is particularly valuable in safety-critical UAM applications where each subsystem's behavior must be auditable.

As an alternative to this modular paradigm, end-to-end deep learning frameworks commonly seek to collapse the entire stack into a single learned policy that maps raw sensor streams directly to control commands, and have demonstrated impressive agility in cluttered natural environments~\cite{loquercio2021learning, zhao2025improved}. However, deployment at the scale and determinism required for UAM remains an open problem. Learned policies are sensitive to distributional shift between training and deployment environments, depend heavily on the quality and diversity of training data, and offer limited formal guarantees on safety-critical constraint satisfaction~\cite{sheltami2026UAV, wang2021survey, tang2025deep}. For this reason, learned planners and controllers are not included in our experimental comparison; we focus instead on the modular architecture with dynamics-aware trajectory optimization methods whose constraint behavior can be audited and certified.

\subsection{Path Planning}
Motion planning is traditionally divided into two stages. Path planning (or global planning) generates a sequence of reference waypoints from origin to destination, typically with reduced geometric fidelity and without accounting for vehicle dynamics. Classical approaches include graph search (A\textsuperscript{*}), sampling-based methods (PRM, RRT, dubins+RRT*), and reactive potential fields (APF), with recent learning-based variants combining graph neural networks and reinforcement learning for cooperative routing~\cite{sheltami2026UAV, zhou2026semanticaware, wang2026dubinsrrt}. In the UAV context, path planning has progressively shifted from pure shortest-path formulations toward multi-objective problems that explicitly reason about energy, noise, privacy, and third-party risk~\cite{pang2022uav, hu2020risk, scott2025Noise, manyam2022Path, zhu2026safetyaware, rivera2026citywide}. Our framework consumes the output of such a global planner as a reference and is agnostic both to the specific algorithm used and to whether that reference is precomputed offline or streamed from a low-frequency online planner. The trajectory optimization stage itself runs online, ingesting the current reference at every replanning cycle.

\subsection{Trajectory Planning}
Trajectory planning (also called local planning or trajectory optimization) is the stage of motion planning that converts a global reference into a dynamically feasible, time-parameterized 4D trajectory over a short horizon, incorporating, to varying degrees, vehicle dynamics, actuator limits, operational restrictions, and obstacle constraints. Dynamics-aware implementations have converged on three dominant real-time algorithmic families: Sequential Quadratic Programming, Differential Dynamic Programming (DDP), and its first-order variant, the iterative Linear Quadratic Regulator (iLQR)~\cite{gill2012sequential, mayne1966secondorder, li2004iterative}. Modern implementations build on these foundations, incorporating Augmented Lagrangian methods for constraint handling (e.g., ALTRO) and hybrid approaches, such as fusing DDP-style closed-loop rollouts into SQP iterates~\cite{howell2019altro, singh2022Optimizinga}. In the UAV domain, optimization-based trajectory planners have demonstrated dynamic obstacle avoidance for small systems in confined spaces, while a parallel line of UAM-scale work has addressed trajectory generation from a UTM airspace-design perspective~\cite{liu2017planning, lindqvist2020nonlinear, oshinetal2024differentiable, liang2024guidance, elsayed2024robust}. ElSayed and Mohamed~\cite{elsayed2024robust}, for example, propose a digital-twin Skyroutes discretization that carves the urban airspace into a network of lanes and generates cubic spline trajectories along them, a networking-style UTM morphology. Their work is complementary to the stratification model targeted here for UAM online trajectory planning.

A practical limitation of conventional SQP, iLQR, and DDP with respect to UAM is that each is typically formulated with a fixed, small set of obstacles encoded directly in the cost or constraint structure~\cite{oshinetal2024differentiable, howell2019altro}. This assumption breaks down in realistic urban settings, where an $800$ to $1000$\,m operational radius may contain several hundred individually relevant buildings.
Furthermore, conventional iLQR and DDP struggle to scale reliably to full multi-state complex dynamics in dense geometry, while many existing SQP-based formulations typically treat obstacle avoidance as an offline preprocessing step rather than an intrinsic constraint inside the optimizer~\cite{mastalli2022feasibility}. To address these gaps, our work reformulates trajectory optimization to reason about complex geometry and vehicle dynamics jointly, utilizing a spatial decomposition approach that keeps computational overhead strictly bounded even as the operational environment scales to city-wide dimensions.

\section{Methodology}\label{sec:ltp}

\begin{figure}[!bt]
    \centering
    \includegraphics[width=0.75\columnwidth]{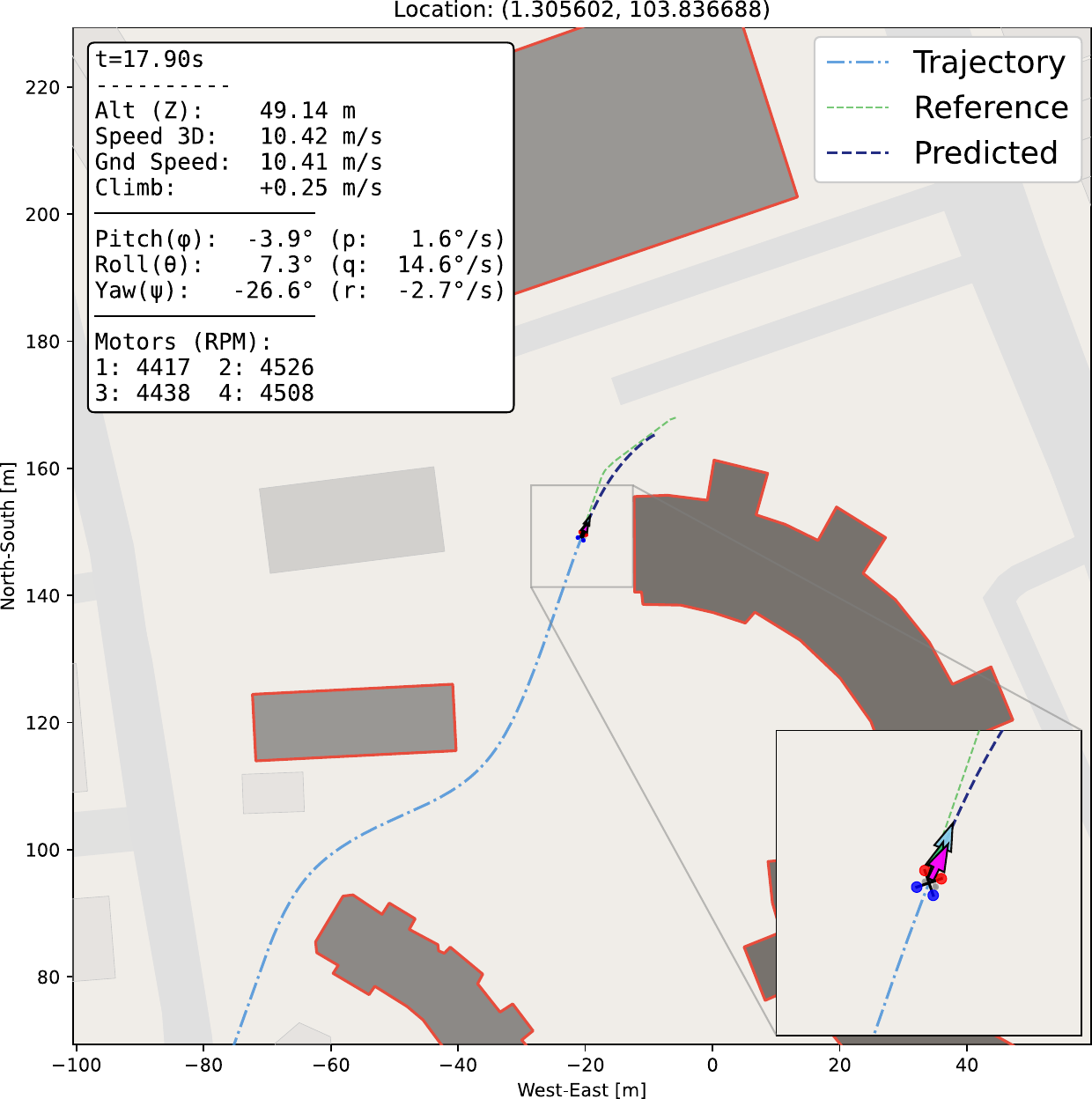}
    \caption{Real time 4D trajectory optimization in ultra low-altitude urban airspace.}\label{fig:gr-route-demo}
\end{figure}

\begin{figure*}[htb]
	\centering
	\includegraphics[width=0.65\textwidth]{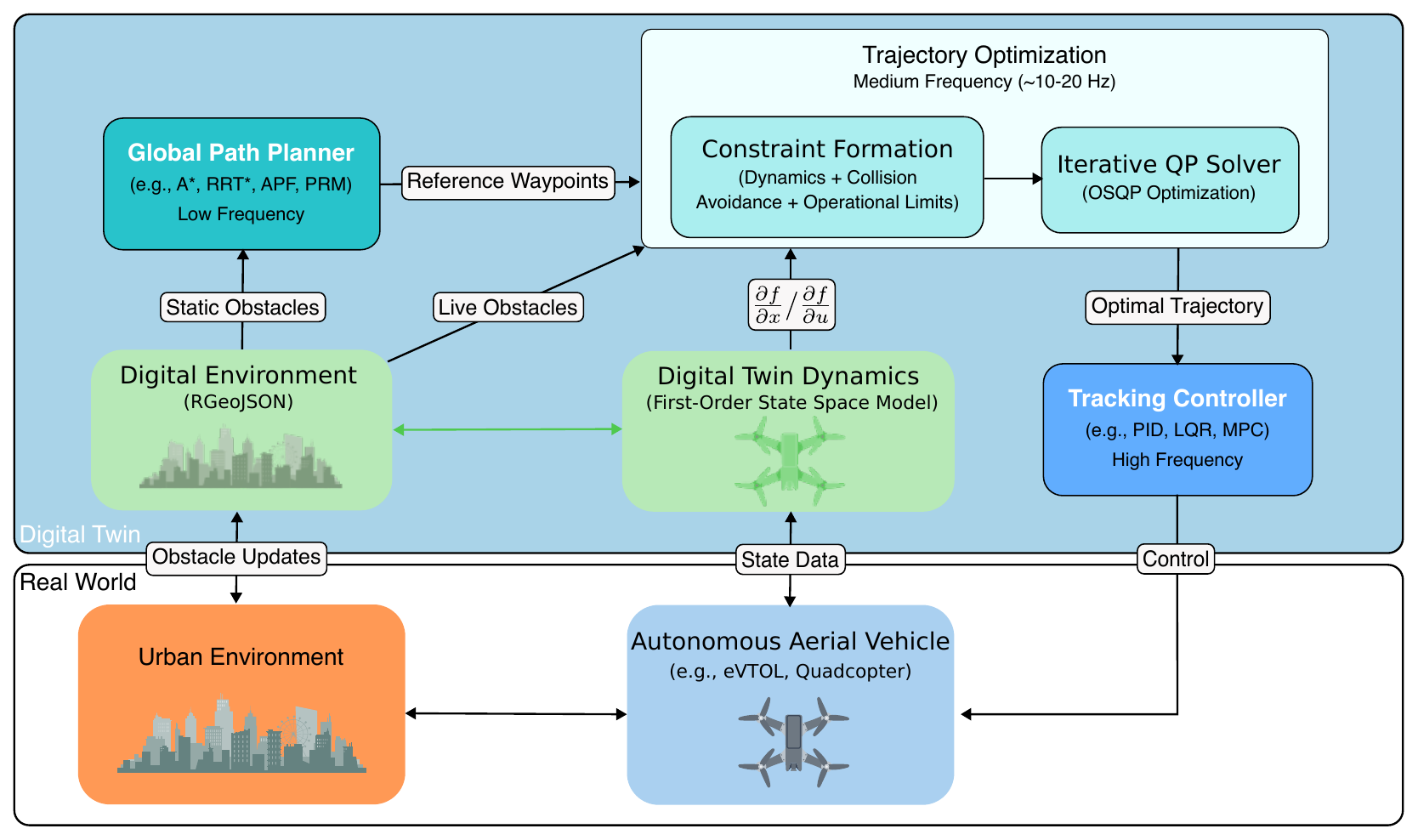}
	\caption{Last-Mile Trajectory Planning (LTP) framework for urban air mobility. Vector skyline and quadcopter adapted from Freepik~\cite{freepikSkylineRawpixel2026, freepikDronesFreepikto2026}.}\label{fig:flowchart}
\end{figure*}

The Last-Mile Trajectory Planning (LTP) framework is an online, dynamics-aware trajectory optimization system that converts a coarse global reference into a feasible, collision-free, and dynamically consistent 4D trajectory for urban aerial mobility (Figure~\ref{fig:gr-route-demo}). As illustrated in Figure~\ref{fig:flowchart}, the framework comprises three tightly coupled components: (1) a \emph{digital-twin UAV model} used for forward state prediction, Jacobian linearization, and state discretization, (2) a \emph{digital environment} that tracks both static infrastructure and obstacles within the UTM channel, and (3) an \emph{online SQP optimizer} that converts the discretized dynamics, operational limits, and live geometric half-planes into a series of Quadratic Programs (QPs) to solve for real-time optimal trajectory.

\subsection{Digital Twin Dynamics Modeling and Discretization}
We define the digital twin of a UAV as a nonlinear, continuous-time, first-order state-space model:
\begin{equation}
    \dot{x}(t) = f(x(t), u(t))
\end{equation}
where $x \in \mathbb{R}^{n}$ and $u \in \mathbb{R}^{m}$ represent the state and control vectors, respectively.

To leverage optimization techniques, we linearize the dynamics around each state and control in a nominal trajectory $(\bar{x}_k, \bar{u}_k)$ derived from either the previous iteration or the current candidate solution. Applying a first-order Taylor expansion, we obtain:
\begin{equation}
\begin{aligned}
    f(x, u) \approx f(\bar{x}_k, \bar{u}_k) &+ \nabla_x f\big|_{\bar{x}_k, \bar{u}_k} (x - \bar{x}_k) \\
                                 &+ \nabla_u f\big|_{\bar{x}_k, \bar{u}_k} (u - \bar{u}_k)
\end{aligned}
\end{equation}
where $A_c\!:=\!\nabla_x f$ and $B_c\!:=\!\nabla_u f$ are the continuous-time Jacobian matrices obtained via automatic differentiation~\cite{bradbury2018jax}. The linearized dynamics can then be written in affine form:
\begin{equation}
    \dot{x}(t) = A_c x(t) + B_c u(t) + d_c
\end{equation}
where the affine drift term is $d_c = f(\bar{x}_k, \bar{u}_k) - A_c \bar{x}_k - B_c \bar{u}_k$.

\subsubsection{System Discretization}
To integrate the continuous-time system over a prediction step $\Delta t$, we employ one of two discretization strategies depending on the complexity of the dynamics and the precision required. 

For simple systems, or when numerical stability and precision are paramount, we apply an exact Zero-Order Hold (ZOH) discretization. First, we construct an augmented matrix $\mathcal{M} \in \mathbb{R}^{(n+m+1) \times (n+m+1)}$:
\begin{equation}
    \mathcal{M} = \begin{bmatrix}
    A_c & B_c & d_c \\
    0 & 0 & 0
    \end{bmatrix}
\end{equation}

Computing the matrix exponential $\Lambda = e^{\mathcal{M} \Delta t}$ yields the exact discrete-time system matrices:
\begin{equation}
    \Lambda = \begin{bmatrix}
    A_d & B_d & g_d \\
    0 & I & 0 \\
    0 & 0 & 1
    \end{bmatrix}
\end{equation}

Conversely, for higher-dimensional or computationally complex systems where matrix exponentiation becomes a bottleneck, we utilize a Forward Euler approximation. This trades integration precision for faster execution speeds:
\begin{align}
    A_d &= I + A_c \Delta t \\
    B_d &= B_c \Delta t \\
    g_d &= d_c \Delta t
\end{align}

Regardless of the chosen strategy, the method results in the discrete affine update law used in our equality constraints:
\begin{equation}
    x_{k+1} = A_d x_k + B_d u_k + g_d
\end{equation}
where $x_k \in \mathbb{R}^n$ and $u_k \in \mathbb{R}^m$ represent the discrete-time state and control variables at step $k$.

By leveraging JAX's JIT compilation, the batched computation of the Jacobians $(A_c, B_c)$ and the selected discretization routine across all time steps in the nominal trajectory are fused into an optimized XLA kernel. This allows for real-time linearization and discretization (typically taking under one millisecond for most evaluated systems) at every inner QP solver iteration~\cite{bradbury2018jax}. Future work will explore integration with differentiable simulators such as Mujoco XLA(MJX) to support non-analytically defined custom digital twin models.

\subsection{Digital Environment and Constraints}
The trajectory optimization is subject to various constraints derived from the digital environment and twin model. We categorize them into three groups: (i) geometric environmental constraints representing static infrastructure, encoded via our RGeoJSON format; (ii) box and linear constraints that enforce operational limits; and (iii) dynamics equality constraints (introduced above) that ensure dynamic feasibility.

\subsubsection{Geometric Environmental Constraints}\label{ssec:halfplanes}
Because urban geometry is non-convex, we perform local convex decomposition along the nominal trajectory. For each digital-twin nominal state $\bar{x}_k$ and set of obstacle geometries $\mathcal{O}$, we identify the closest surface points $p_{obs} \in \mathcal{O}$ via linear projection and compute the repulsion vector $v_{rep} = x_{pos} - p_{obs}$. We then formulate a separating hyperplane defined by the unit normal $n = v_{rep} / \|v_{rep}\|$ for each face. To ensure the vehicle maintains a safety distance of at least $d_{safe}$, we enforce:
\begin{equation}
    n^\top x_{pos} \geq n^\top p_{obs} + d_{safe}
\end{equation}

Across all nominal states $\bar{x}_k$, these inequalities are aggregated into the half-plane matrix $\bar{A}_{env}$ and bound vector $\bar{b}_{env}$:
\begin{equation}
    \bar{A}_{env, k}\, x_k \leq \bar{b}_{env, k}
\end{equation}

\begin{figure}[!tb]
    \centering
    \begin{subfigure}[b]{0.48\columnwidth}
        \centering
        \includegraphics[width=\columnwidth]{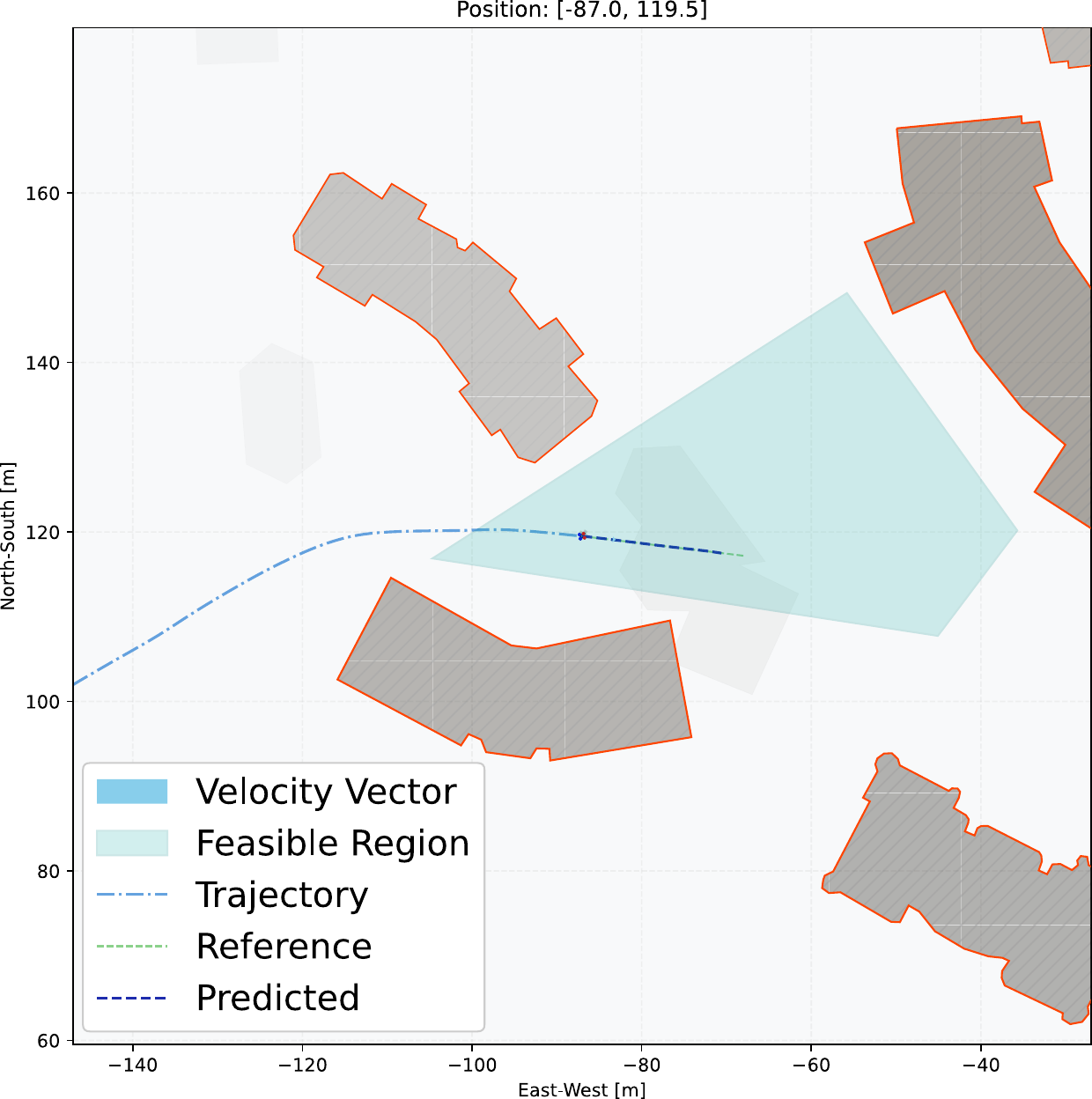}
        \caption{Relaxed boundary.}\label{fig:first}
    \end{subfigure}
    \hfill
    \begin{subfigure}[b]{0.48\columnwidth}
        \centering
        \includegraphics[width=\columnwidth]{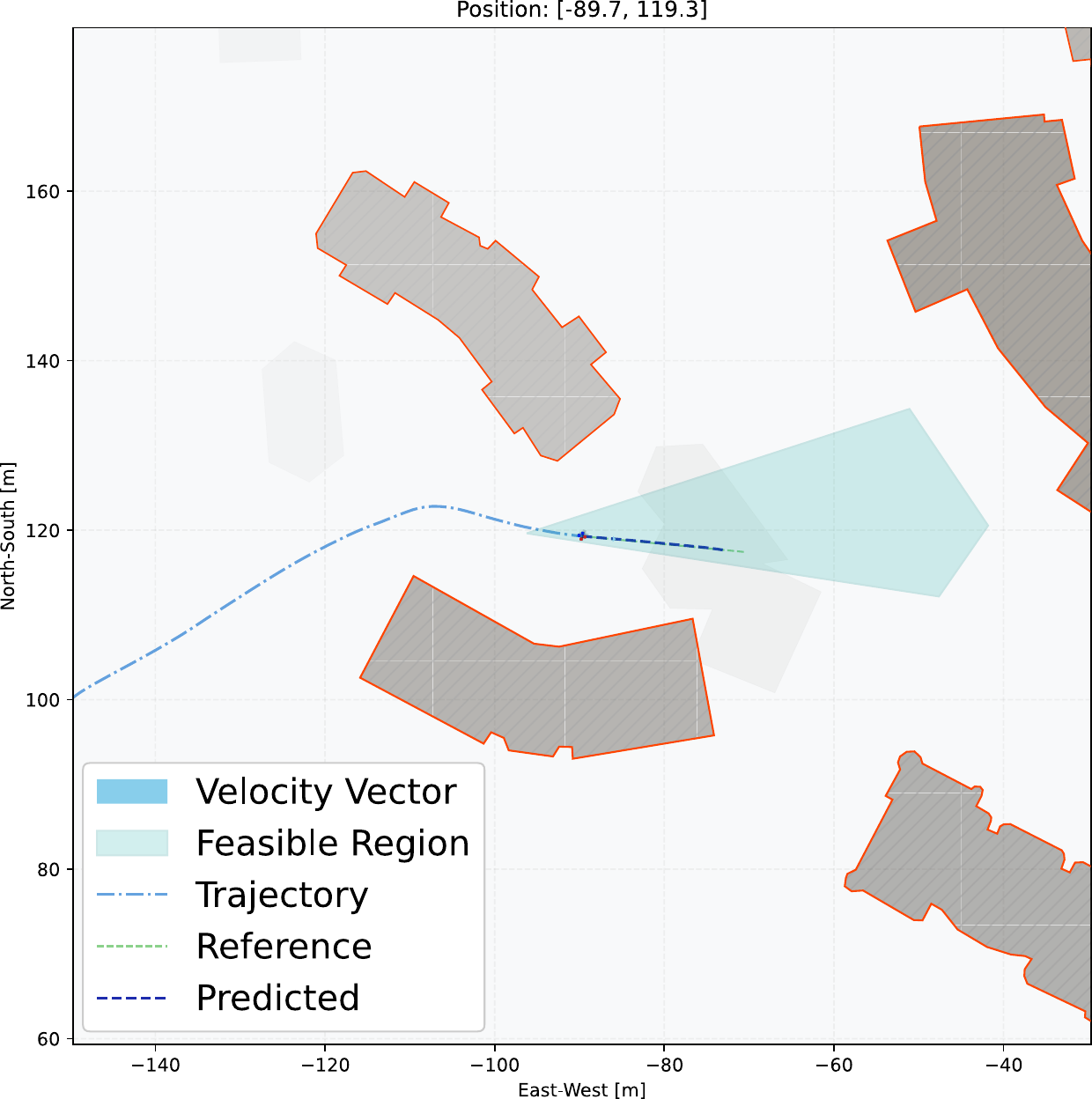}
        \caption{Strict boundary.}\label{fig:second}
    \end{subfigure}
    \caption{Geometric constraint feasible space (highlighted in sky blue) shrinks as the safety distance boundary $d_{safe}$ from buildings is increased with respect to the current state.}\label{fig:boundary-configs}
\end{figure}

The safety distance $d_{safe}$ is exposed as a tunable parameter, allowing the operator to control how close the vehicle is permitted to approach any infrastructure surface. This relationship is illustrated in Figure~\ref{fig:boundary-configs}. Each polygon edge in the highlighted feasible space is a separating hyperplane, offset from the nearest buildings face by $d_{safe}$, and the feasible space itself is the intersection of these half-planes with respect to the current state. Increasing $d_{safe}$ pushes every bounding line inward, progressively shrinking the feasible region.

\begin{figure}[tb]
    \centering
    \includegraphics[width=0.61\columnwidth]{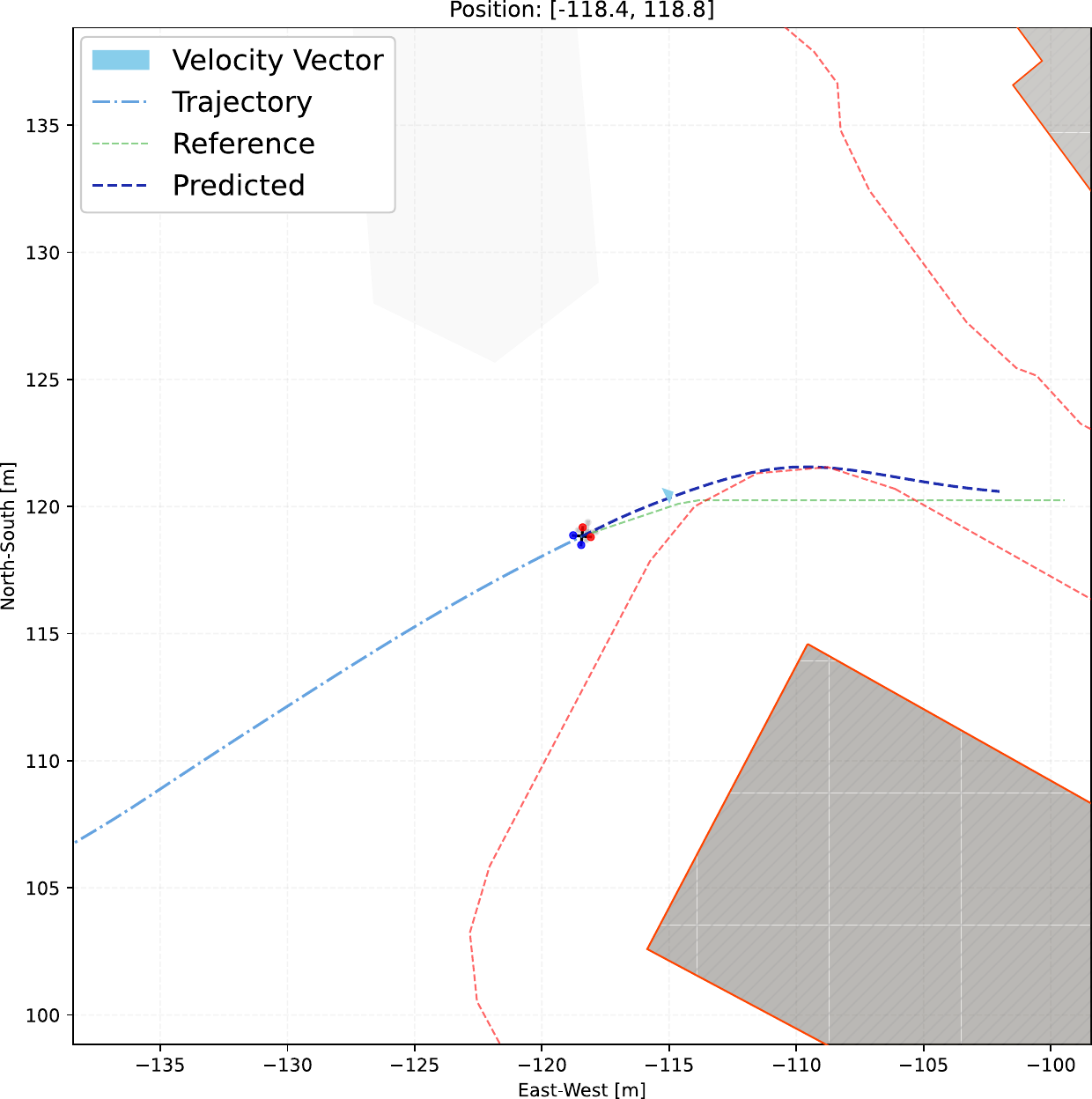}
    \caption{Planned trajectory deviation from an ill-conditioned reference when approaching a building corner.}\label{fig:boundary}
\end{figure}

Because the half-planes are regenerated from the nominal trajectory at every inner solver call, the formulation also confers reference robustness: when the global reference passes too close to (or through) an obstacle, the active half-planes pull the predicted states back into the feasible region, so the planner deviates autonomously from the unsafe reference rather than tracking it into a collision or violation (Figure~\ref{fig:boundary}).

\subsubsection{Variable-Scale Quadtree Decomposition}
\begin{figure}[htb]
    \centering
    \includegraphics[width=0.62\columnwidth]{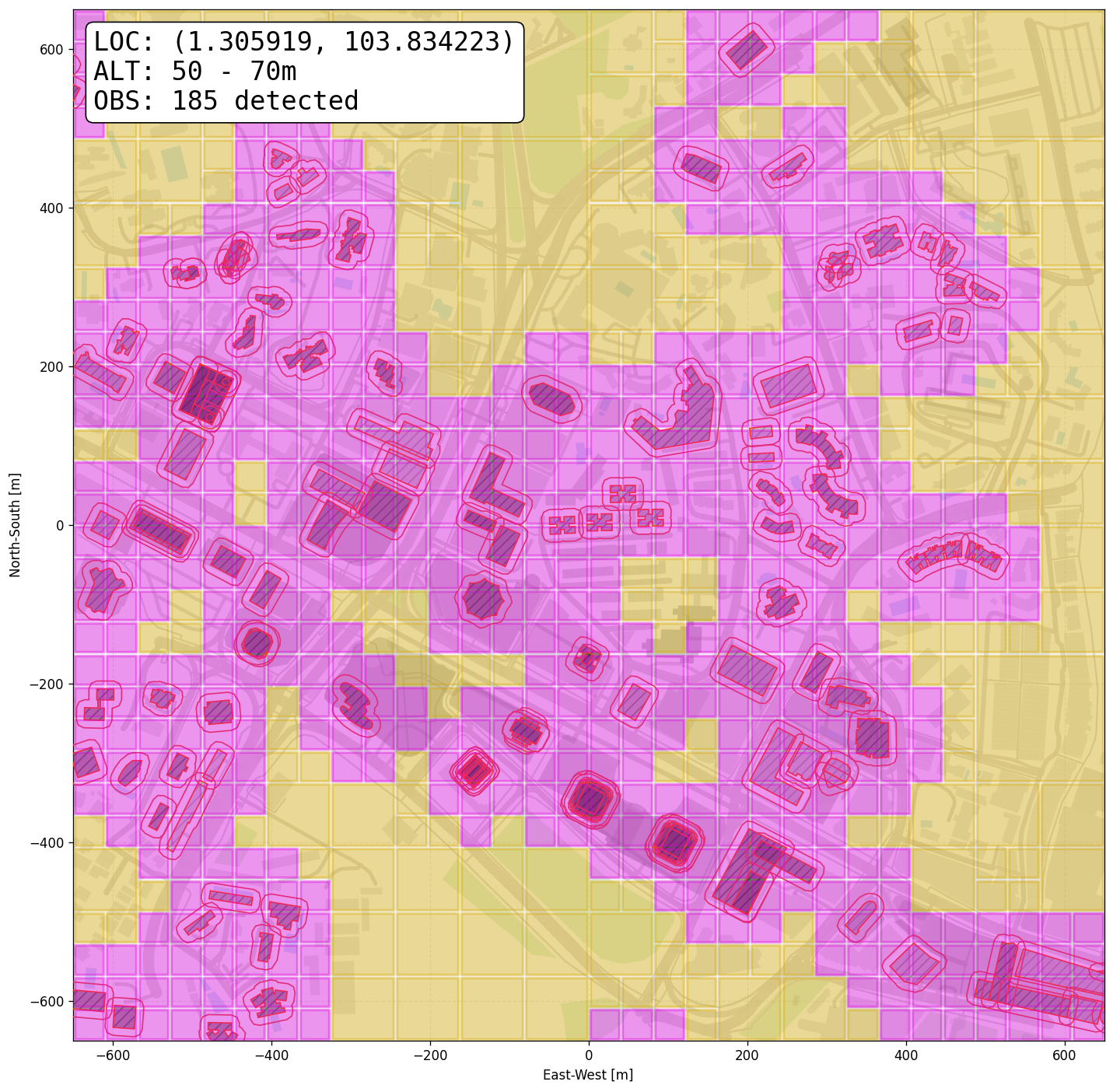}
    \caption{Quadtree decomposition of the Orchard Road UAM channel into sectors, each maintaining a local registry of live obstacles.}\label{fig:quad-decomp}
\end{figure}

As the operational area grows to city-scale dimensions, considering all obstacles in a sector becomes computationally expensive: a typical UAM channel of $800$ to $1000$\,m radius can contain several hundred individually relevant buildings, and naively appending one half-plane per building face per horizon step inflates the QP beyond what can be solved at a planning rate. To preserve real-time performance while retaining global awareness, we partition the airspace channel using a variable-scale quadtree decomposition (Fig.~\ref{fig:quad-decomp}). Each leaf cell carries an obstacle registry listing the obstacles whose influence intersect that cell.

At each inner solver call in the trajectory replanning cycle, the planner queries the registries of the cells that the nominal trajectory horizon traverses and constructs half-planes only for the locally relevant subset; obstacles outside the current sector contribute zero rows to $A_{env}$ and are therefore invisible to OSQP\@. SQP is well suited to this regime because the active half-plane set varies smoothly between calls, allowing OSQP's warm-starting to amortize most of the factorization cost across solves. The net effect is that the per-call QP size, and therefore solve time, depends on a small local obstacle density rather than on the global map size, which is the property that lets the framework scale from a single block to an entire downtown without re-tuning. Furthermore, the variable-scale nature of the quadtree decomposition mitigates the computational impact of exceptionally dense local obstacle clusters.

\subsubsection{Operational Constraints}\label{ssec:oplimits}
Operational constraints encompass any physical, regulatory, or mission-specific restrictions that can be expressed as box bounds or linear inequalities over the decision variables. 

Basic kinematic limits and actuator saturations are enforced natively as box constraints on the state and control variables:
\begin{align*}
    u_{min} \leq u_k \leq u_{max}, \qquad
    x_{min} \leq x_k \leq x_{max}
\end{align*}
where $x_{min/max}$ encodes bounds such as velocity limits, generalized orientation constraints, and altitude floors or ceilings, while $u_{min/max}$ captures the specific control input limits of the UAV.

Beyond simple box bounds, the framework accommodates any generalized operational requirement formulated as a linear inequality. Relevant examples include:
\begin{itemize}
    \item \textbf{Control-rate limits:} $\;\lvert u_{k+1} - u_k \rvert \leq \Delta u_{max}$, which suppress aggressive control commands and reduce wear on the actuators.
    \item \textbf{Virtual geofencing:} half-space constraints $a^\top x_{pos} \leq b$ that cordon off no-fly zones, restricted airspace, or terminal-area corridors not represented in the physical building geometry.
    \item \textbf{Bounding-state corridors:} time-varying box constraints $\underline{x}_k \leq x_k \leq \overline{x}_k$ that confine a portion of the trajectory to a tube around the reference, useful for departure or approach segments where conformance is required.
    \item \textbf{Coupled state-control limits:} linear couplings that enforce combined restrictions, written in the general form $A_{lin}\, z \leq b_{lin}$.
\end{itemize}

All such operational constraints are absorbed into the same $A_{con}$ and $(l, u)$ structure used for the dynamics and environment geometry.

\subsection{Quadratic Programming Trajectory Formulation}
We formulate trajectory generation using Sequential Quadratic Programming (SQP). At each SQP iteration, we solve the following convex quadratic optimization problem over an $N$-step horizon until convergence:
\begin{equation}
    \begin{aligned}
        \min_{x, u} \quad & \sum_{k=1}^{N} \| x_k - x_k^{ref} \|_{Q}^2 + \sum_{k=0}^{N-1} \| u_k \|_{R}^2 + \| x_N - x_N^{ref} \|_{Q_f}^2 \\
        \text{s.t.} \quad & x_{k+1} = A_{d,k} x_k + B_{d,k} u_k + g_{d,k} \\
        & \bar{A}_{env,k}\, x_k \leq \bar{b}_{env,k} \\
        & x_{min} \leq x_k \leq x_{max}, \quad u_{min} \leq u_k \leq u_{max} \\
        & A_{lin}\, z \leq b_{lin} \quad \text{(optional state-control limits)} \\
        & x_0 = x_{init}
    \end{aligned}
\end{equation}

To solve the problem efficiently with the OSQP solver, we transform it into the standard QP form:
\begin{equation}
    \begin{aligned}
    & \underset{z}{\text{minimize}} & & \tfrac{1}{2} z^\top P z + q^\top z \\
    & \text{subject to} & & l \leq A_{con}\, z \leq u
    \end{aligned}
\end{equation}
where the decision vector $z \in \mathbb{R}^{N(n+m)}$ is constructed by stacking the predicted states followed by the control sequence. Given that the initial state $x_0$ is a fixed parameter, the decision variables are defined as:
\begin{equation}
    z = \begin{bmatrix} x_1^\top & \cdots & x_N^\top & u_0^\top & \cdots & u_{N-1}^\top \end{bmatrix}^\top
\end{equation}

Expanding the quadratic tracking error $(x - x_{ref})^\top Q (x - x_{ref}) = x^\top Q x - 2 x_{ref}^\top Q x + C$, we derive the block-diagonal Hessian $P$ and linear cost vector $q$:
\begin{equation}
    P = \text{diag}\left( 2Q, \dots, 2Q_f, 2R, \dots, 2R \right)
\end{equation}
\begin{equation}
    q = \begin{bmatrix} -2Q x_1^{ref} \\ \vdots \\ -2Q_f x_N^{ref} \\ 0_{Nm \times 1} \end{bmatrix}
\end{equation}

The constraint matrix $A_{con}$ is a sparse vertical stack of dynamics equalities, geometric half-plane inequalities, linear constraints, and variable bounds.

For the dynamics $x_{k+1} = A_{d,k} x_k + B_{d,k} u_k + g_{d,k}$, we rearrange terms to isolate the variables in $z$:
\begin{equation}
    \underbrace{I}_{\text{coeff of } x_{k+1}} x_{k+1} \underbrace{- A_{d,k}}_{\text{coeff of } x_k} x_k \underbrace{- B_{d,k}}_{\text{coeff of } u_k} u_k = g_{d,k}
\end{equation}
This yields the sparse dynamics block $A_{dyn} \in \mathbb{R}^{Nn \times N(n+m)}$:
\begin{equation}
    A_{dyn} = \begin{bmatrix}
    I & 0 & \cdots & 0 & -B_{d,0} & 0 & \cdots \\
    -A_{d,1} & I & \cdots & 0 & 0 & -B_{d,1} & \cdots \\
    0 & -A_{d,2} & \ddots & 0 & 0 & 0 & \ddots \\
    \vdots & \vdots & \cdots & I & \vdots & \vdots & \cdots
    \end{bmatrix}
\end{equation}
The bounds vectors $l_{dyn}$ and $u_{dyn}$ for these rows enforce the equality constraint:
\begin{equation}
    l_{dyn} = u_{dyn} = \begin{bmatrix} A_{d,0} x_{0} + g_{d,0} \\ g_{d,1} \\ \vdots \\ g_{d,N-1} \end{bmatrix}
\end{equation}
Note that for the first step ($k=0$), the term $A_{d,0} x_0$ is moved to the right-hand side bound vector, since $x_0$ is fixed.

The final constraint system is assembled as:
\begin{equation}
    A_{con} = \begin{bmatrix} A_{dyn} \\ A_{env} \\ A_{lin} \\ I \end{bmatrix}, \quad
    l = \begin{bmatrix} l_{dyn} \\ -\infty \\ -\infty \\ z_{min} \end{bmatrix}, \quad
    u = \begin{bmatrix} u_{dyn} \\ b_{env} \\ b_{lin} \\ z_{max} \end{bmatrix}
\end{equation}
where $A_{env}$ contains the half-plane rows aligned with the corresponding state indices in $z$, $A_{lin}$ aggregates any linear constraints over the decision variables, and the trailing identity block enforces the box bounds. Between successive cycles, only the rows that depend on the current operating point ($A_{dyn}$, $A_{env}$, and the affine bounds) need to be updated. The structural sparsity pattern is preserved, so OSQP can warm-start from the previous solution and amortize most of the factorization cost across solves.

\section{Experiments and Results}\label{sec:results}

To evaluate the efficacy of the proposed LTP framework, we conducted comprehensive cargo delivery simulation experiments to quantify its performance and analyzed case studies to demonstrate its practical utility.

\subsection{Experimental Setup}

\subsubsection{Digital Twin}
Simulations are built around a 6-DOF Newton--Euler cargo quadcopter digital twin model~\cite{islam2017dynamics}.
The 12-dimensional state and 4-dimensional control vectors are
\begin{equation}
\begin{aligned}
    x &= \bigl[p^\top,\; v^\top,\; \xi^\top,\; \omega^\top\bigr]^\top \in \mathbb{R}^{12}\\
    u &= \bigl[w_1,\; w_2,\; w_3,\; w_4\bigr]^\top \in \mathbb{R}^4,
\end{aligned}
\label{eq:state}
\end{equation}
where $p\!=\![x,y,z]^\top$ is the world-frame position, $v\!=\![v_x,v_y,v_z]^\top$ the linear velocity,
$\xi\!=\![\phi,\theta,\psi]^\top$ the roll--pitch--yaw Euler angles, $\omega\!=\![p,q,r]^\top$ the body
angular rates, and $w_i$ the individual rotor speeds in an $\times$-frame layout.

The four rotor speeds are mapped to collective thrust and body torques through an actuator mixer,
\begin{equation}
  \begin{bmatrix} u_1 \\ \tau_\phi \\ \tau_\theta \\ \tau_\psi \end{bmatrix}
  = \underbrace{\begin{bmatrix}
      k_f                & k_f                & k_f                & k_f                \\[2pt]
      -\tfrac{k_f l}{\sqrt{2}} & -\tfrac{k_f l}{\sqrt{2}} & \tfrac{k_f l}{\sqrt{2}} & \tfrac{k_f l}{\sqrt{2}} \\[2pt]
      -\tfrac{k_f l}{\sqrt{2}} &  \tfrac{k_f l}{\sqrt{2}} & \tfrac{k_f l}{\sqrt{2}} & -\tfrac{k_f l}{\sqrt{2}} \\[2pt]
       k_M              & -k_M               &  k_M               & -k_M
  \end{bmatrix}}_{\mathbf{M}}
  \begin{bmatrix} w_1^2 \\ w_2^2 \\ w_3^2 \\ w_4^2 \end{bmatrix}
  \label{eq:mixer}
\end{equation}
where $u_1$ is the collective thrust, $\tau_b\!=\![\tau_\phi,\tau_\theta,\tau_\psi]^\top$ are the roll,
pitch, and yaw body torques. Motors 1--\,4 are positioned at $-45^\circ, -135^\circ, +135^\circ,
+45^\circ$ relative to the forward axis, alternating CW/CCW.

The continuous-time state-space model $\dot{x}(t)=f(x,u)$ is described by:
\begin{equation}
\begin{aligned}
  \dot{x}(t) &=
  \begin{bmatrix} \dot{p} \\[6pt] \dot{v} \\[6pt] \dot{\xi} \\[6pt] \dot{\omega} \end{bmatrix}
  =
  \begin{bmatrix}
    v \\[6pt]
    \dfrac{u_1}{m}\,R(\xi)\,e_3
      - \dfrac{k_t}{m}\,v
      - g\,e_3 \\[10pt]
    \mathbf{T}(\xi)\,\omega \\[6pt]
    \begin{bmatrix}
      I_x^{-1}\!\left(\tau_\phi   + (I_y\!-\!I_z)\,qr - I_r q\,\Omega_r - k_r p\right)\\[2pt]
      I_y^{-1}\!\left(\tau_\theta + (I_z\!-\!I_x)\,pr + I_r p\,\Omega_r - k_r q\right)\\[2pt]
      I_z^{-1}\!\left(\tau_\psi   + (I_x\!-\!I_y)\,pq                   - k_r r\right)
    \end{bmatrix}
  \end{bmatrix}
\end{aligned}
  \label{eq:dynamics}
\end{equation}
where, letting $c_\alpha\!:=\!\cos\alpha$, $s_\alpha\!:=\!\sin\alpha$, and
$t_\theta\!:=\!\tan\theta$, the body $z$-axis in the world frame and the
Euler-angle kinematic mapping are defined as
\begin{equation*}
  R(\xi)\,e_3 =
  \begin{bmatrix}
    s_\phi s_\psi + c_\phi c_\psi s_\theta \\
    c_\phi s_\psi s_\theta - c_\psi s_\phi \\
    c_\phi c_\theta
  \end{bmatrix},
  \quad
  \mathbf{T}(\xi) =
  \begin{bmatrix}
    1 & s_\phi\,t_\theta & c_\phi\,t_\theta \\
    0 & c_\phi           & -s_\phi          \\
    0 & s_\phi/c_\theta  & c_\phi/c_\theta
  \end{bmatrix},
\end{equation*}
$e_3\!=\![0,0,1]^\top$, $\Omega_r\!=\!w_1\!-\!w_2\!+\!w_3\!-\!w_4$ is the net signed rotor angular
momentum (gyroscopic term), and $I\!=\!\mathrm{diag}(I_x,I_y,I_z)$.
Physical parameters are listed in Table~\ref{tab:quad_physical}.

\begin{table}[tb]
    \centering
    \footnotesize
    \caption{Quadcopter Physical \& Aerodynamic Parameters}\label{tab:quad_physical}
    \setlength{\tabcolsep}{4pt}
    \renewcommand{\arraystretch}{1.1}
    \begin{tabular*}{\columnwidth}{@{\extracolsep{\fill}} lll}
        \toprule
        \textbf{Sym.} & \textbf{Parameter} & \textbf{Value [Unit]} \\
        \midrule
        $m$   & Mass          & $5.2$ [kg] \\
        $l$   & Arm length    & $0.35$ [m] \\
        $I$   & Inertia       & $\text{diag}(.04,.04,.08)$ \\
        $I_r$ & Motor inertia & $2.0\!\cdot\!10^{-4}$ [$\text{kg}\!\cdot\!\text{m}^2$] \\
        $g$   & Gravity       & $9.807$ [$\text{m}/\text{s}^2$] \\
        $k_f$ & Thrust coeff. & $6.0\!\cdot\!10^{-5}$ [$\text{N}\!\cdot\!\text{s}^2$] \\
        $k_M$ & Moment coeff. & $1.0\!\cdot\!10^{-6}$ [$\text{N}\!\cdot\!\text{m}\!\cdot\!\text{s}^2$] \\
        $k_t$ & Trans.\ drag  & $\text{diag}(0.5,0.5,0.8)$ [$\text{N}\!\cdot\!\text{s}/\text{m}$] \\
        $k_r$ & Rot.\ drag    & $0.2\cdot\mathbf{I}_{3}$ [$\text{N}\!\cdot\!\text{m}\!\cdot\!\text{s}$] \\
        \bottomrule
    \end{tabular*}
\end{table}

\subsubsection{Urban Environments}
To evaluate generalization across diverse urban morphologies, experiments were conducted in five real-world environments extracted from OpenStreetMap\cite{openstreetmap2017planet}:
\emph{Austin, Texas},
\emph{Boston, Massachusetts},
\emph{Orchard Road district, Singapore},
\emph{Marina Bay district, Singapore}, and
\emph{Hong Kong}.
Building footprints for each environment are ingested as RGeoJSON polygons and converted to convex half-plane constraints at runtime.
The five environments span a broad spectrum of urban density, building height, and street-grid regularity (from the low-rise grid of Austin to the ultra-dense high-rise cluster of Hong Kong), providing a stringent and diverse evaluation of the framework.
Per-environment area radii and UTM cruise altitudes are reported alongside route counts in Table~\ref{tab:environments}.

\begin{table}[tb]
    \centering
    \footnotesize
    \caption{Urban Environments}\label{tab:environments}
    \setlength{\tabcolsep}{4pt}
    \renewcommand{\arraystretch}{1.1}
    \begin{tabular*}{\columnwidth}{@{\extracolsep{\fill}} llll}
        \toprule
        \textbf{Environment} & \textbf{Radius [m]} & \textbf{Alt. Channel\ [m]} & \textbf{Routes} \\
        \midrule
        Austin, TX        & 700  & $50-70$  & 10 \\
        Boston, MA        & 500  & $40-60$  & 12 \\
        Orchard Road, Singapore    & 1000 & $50-70$  & 14 \\
        Marina Bay, Singapore      & 600  & $50-70$  & 10 \\
        Hong Kong         & 700  & $90-100$ & 11 \\
        \midrule
        \textbf{Total}             &      &    & \textbf{57} \\
        \bottomrule
    \end{tabular*}
\end{table}

Operational limits and trajectory planner parameters are summarized in Table~\ref{tab:quad_limits}.
Rotor speeds are bounded to $[100,\;1.5\,w_\text{hover}]$\,rad/s to prevent stall and actuator saturation, where $w_\text{hover}=\sqrt{mg/(4k_f)}\approx 464$\,rad/s.
Box constraints on the state enforce the velocity, tilt, and altitude limits shown in Table~\ref{tab:quad_limits}. No cost on the control is applied.
Geometry constraints (half-plane rows in $A_\text{con}$) are rebuilt at every inner solver call using the current nominal trajectory.

\begin{table}[tb]
    \centering
    \footnotesize
    \caption{Quadcopter Operational Limits and Simulation Parameters}\label{tab:quad_limits}
    \setlength{\tabcolsep}{4pt}
    \renewcommand{\arraystretch}{1.1}
    \begin{tabular*}{\columnwidth}{@{\extracolsep{\fill}} lll}
        \toprule
        \textbf{Sym.} & \textbf{Parameter} & \textbf{Value [Unit]} \\
        \midrule
        $d_{safe}$      & Boundary distance  & $10.0$ [m] \\
        $v_{horz}$      & Max horiz.    & $40.0$ [m/s] \\
        $v_{vert}$      & Vert.\ range  & $[-3.0,\ 4.0]$ [m/s] \\
        $\phi,\,\theta$ & Max tilt      & $0.6$ [rad] \\
        $\omega$        & Max rate      & $2.5$ [rad/s] \\
        $\Delta t$      & Time step     & $0.1$ [s] \\
        $T$             & Horizon       & $2.5$ [s] \\
        $N$             & Horizon steps & $25$ \\
        $Q$             & State cost matrix & $\begin{array}[t]{@{}l@{}}
10^{2}\!\cdot\!\text{diag}(20, 7, 20, 7, 50, 5, \\
\hphantom{10^{2}\!\cdot\!\text{diag}()}10, 10, 10, 50, 50, 10)
\end{array}$ \\
        $R$             & Control cost matrix & $\mathbf{0}_{4\times4}$ \\
        $Q_f$           & Final state cost & $2\!\cdot\!Q$ \\
        \bottomrule
    \end{tabular*}
\end{table}

\subsubsection{Computational Platform}
All simulations were executed on a single workstation equipped with an Intel\textsuperscript{\textregistered} Core\textsuperscript{\texttrademark} i5-14600KF processor (14 cores / 20 threads, 3.5 GHz base clock speed, up to 5.3\,GHz) and 32\,GB of DDR5 RAM\@.
No GPU acceleration was used. JAX was configured to run exclusively on the CPU backend (XLA/LLVM) and OSQP to use its built-in sparse algebra backend.
This setup reflects a realistic onboard or edge-computing deployment scenario, where high-end GPU resources are unavailable.

\subsection{Quantitative Comparative Analysis}

We conducted a benchmark study comprising 1710 cargo delivery flights across all five environments, six planning algorithms, and five target speeds (5, 8, 10, 12, and 15\,m/s).
All reference paths were pre-verified to lie at least $d_\text{safe}$ from any building surface, so any clearance violation is attributable solely to the planner.
Two primary metrics are reported in Table~\ref{tab:benchmark_quad}: \emph{Success Rate} (SR), the fraction of flights reaching the goal without numerical divergence, and \emph{Clearance Rate} (CR), the fraction of successful flights that maintained a separation of at least $d_\text{safe}\!-\!0.1$\,m from every obstacle throughout the trajectory. The remaining columns report solve time per planning call, minimum clearance, travel time, path length, and control effort (Ctrl.\ Eff.), defined here as the mean norm of the rotor-speed along the trajectory as a proxy for actuation energy.

\begin{table*}[htb]
    \centering
    \footnotesize
    \caption{UAM Quadcopter Trajectory Planning Benchmark: Aggregate over 1710 flights (57 routes $\times$ 6 algorithms $\times$ 5 speeds, five cities). Metrics marked \textendash{} are undefined due to 0\% SR. All metrics computed on successful trials only.}\label{tab:benchmark_quad}
    \setlength{\tabcolsep}{5pt}
    \renewcommand{\arraystretch}{1.1}
    \begin{tabular*}{\textwidth}{@{\extracolsep{\fill}} lrrrrrrr}
        \toprule
        \textbf{Algorithm} & \textbf{SR (\%)} & \textbf{CR (\%)} & \textbf{Solve (ms)} & \textbf{Min Clear.\ (m)} & \textbf{Travel (s)} & \textbf{Path (m)} & \textbf{Ctrl.\ Eff.} \\
        \midrule
        DDP                              & 0.0   & \textendash & \textendash       & \textendash            & \textendash         & \textendash        & \textendash \\
        iLQR                             & 98.2  & 95.4        & $137.60\pm19.65$  & $11.79\pm20.13$        & $34.90\pm26.56$     & $312.7\pm191.8$    & $932.4\pm18.0$ \\
        SQP                              & 100.0 & 95.4        & $25.04\pm1.10$    & $11.69\pm20.02$        & $35.49\pm26.59$     & $310.2\pm190.8$    & $935.7\pm13.1$ \\
        SQP (Relax.\ Limits)             & 99.6  & 95.1        & $23.98\pm0.36$    & $11.70\pm20.05$        & $35.26\pm26.67$     & $310.5\pm191.0$    & $938.0\pm15.9$ \\
        SQP (Geo.) \textbf{(Ours)}       & 100.0 & 98.9        & $31.56\pm4.27$    & $11.71\pm20.00$        & $35.50\pm26.59$     & $310.2\pm190.8$    & $935.4\pm12.7$ \\
        SQP (Geo.\ + Relax.) \textbf{(Ours)} & 100.0 & \textbf{100.0} & $30.64\pm4.09$ & $11.70\pm20.01$   & $35.20\pm26.65$     & $310.1\pm190.8$    & $938.0\pm15.9$ \\
        \bottomrule
    \end{tabular*}
\end{table*}

\subsubsection{DDP Divergence}
DDP achieved a 0\% success rate on every quadcopter trial.
Its second-order quadratic approximation is ill-conditioned for the 12-state Newton-Euler model at the tested speeds: the highly nonlinear rotational dynamics drive the value-function Hessian to lose positive definiteness within a few iterations, causing backward-pass failures and erroneous control outputs.
This is a known limitation of vanilla DDP on high-dimensional nonlinear systems~\cite{mastalli2022feasibility} and is absent in the WMR sanity check (below).

\subsubsection{iLQR vs.\ SQP}
iLQR recovers an SR of 98.2\% by dropping the second-order state correction, yet achieves only 95.4\% CR and requires $137.60\pm19.65$\,ms per planning call, roughly $5.5\times$ slower than the base SQP planner ($25.04\pm1.10$\,ms).
The gap stems from iLQR's full-horizon forward--backward sweeps at every iteration, whereas SQP reduces each outer iterate to a few sparse OSQP solves.
With near-identical path lengths (${\approx}311$\,m) and travel times (${\approx}35$\,s), trajectory quality is comparable; SQP's advantage is purely computational.

\subsubsection{Effect of Geometric Constraints}
Adding the live half-plane constraints (SQP Geo.) raises CR from 95.4\% to 98.9\% while maintaining solve time competitive for real-time trajectory replanning. The quadtree decomposition limits the number of active half-plane rows to those in the UAV's local sector, which keeps the QP size small regardless of city density.
Combining geometric constraints with relaxed operational limits (SQP Geo.\ + Relax.) achieves a perfect 100\% CR across all 285 flights while maintaining similar $0.03$\,s solve time, demonstrating that the two mechanisms are complementary. Geometry constraints push the trajectory away from obstacles, and limit relaxation provides the additional kinematic slack needed to satisfy both the avoidance and the dynamics constraints simultaneously at high speeds.

We further note that the solve-time deviations remain small across all flights (e.g., $31.56\pm4.27$\,ms for SQP Geo.). The deterministic OSQP iterations, inter-call warm-starting, and quadtree-bounded QP size make per-call timing depend on local obstacle density rather than on route length or city scale. This low variance tightens the planner's real-time tolerance, as worst-case latency stays close to the mean, a desirable property for onboard scheduling and certification. By contrast, the large spread in the clearance and travel-time columns (e.g., $11.71\pm20.00$\,m and $35.50\pm26.59$\,s) confirms that the benchmark itself spans diverse and demanding route geometries.

\subsubsection{Speed vs.\ Clearance}
\begin{figure}[tb]
    \centering
    \includegraphics[width=\columnwidth]{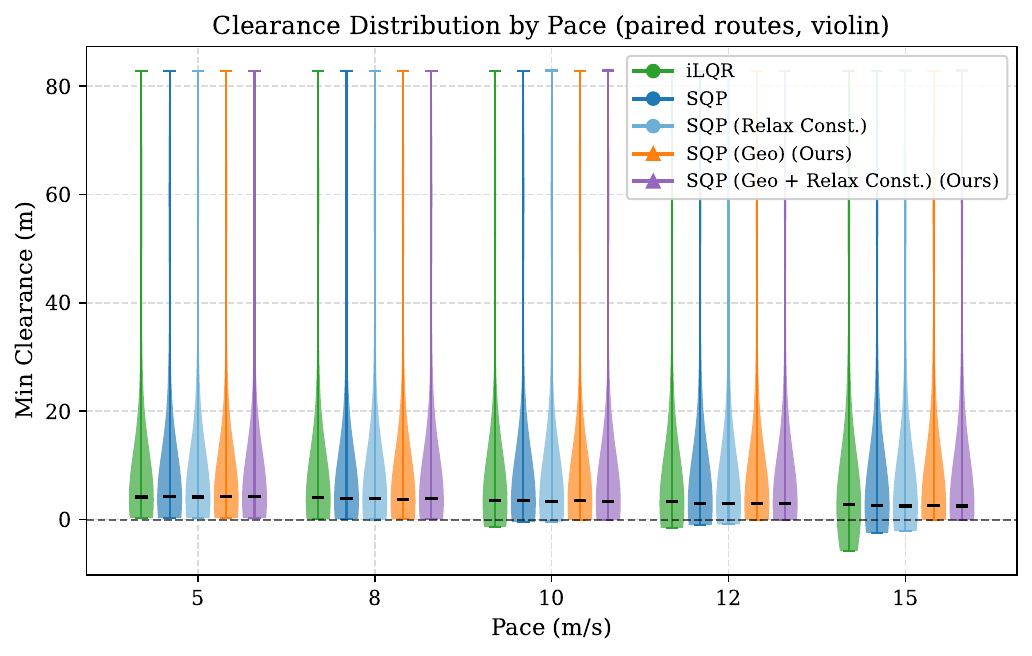}
    \caption{As the UAV is requested to travel at a faster pace, iLQR and conventional SQP struggle to maintain safety distance from buildings.}\label{fig:pace-vs-clr}
\end{figure}

As Figure~\ref{fig:pace-vs-clr} shows, minimum clearance degrades noticeably for iLQR and base SQP at 12--15\,m/s: aggressive dynamic turns cause the UAV to cut corners and challenge the safety margin even when the reference path is clear.
The geometric SQP variants maintain their clearance margin across all speeds, confirming that the half-plane constraints are dominant mechanisms for high-speed safety. 
To account for numerical precision at high speeds, we defined the clearance tolerance as $0.1$\,m below $d_{safe}$.

As a sanity check, we applied all algorithms to a simplified Wheeled Mobile Robot (WMR) scenario with three routes and ten static obstacles, incorporating obstacle avoidance via the Augmented Lagrangian method for DDP and iLQR. As shown in Table~\ref{tab:wmr-summary}, all algorithms, including DDP, achieve $100\%$ SR and CR, suggesting that the quadcopter divergence stems from the system's strongly nonlinear 12-dimensional dynamics and complex urban environment rather than an implementation artifact. All source code and configurations are publicly released, though absolute execution times vary across hardware.

\begin{table}[tb]
    \centering
    \footnotesize
    \caption{WMR Sanity-Check Benchmark (10 obstacles). All algorithms achieve $100\%$ SR and CR in this low-complexity setting.}\label{tab:wmr-summary}
    \setlength{\tabcolsep}{4pt}
    \renewcommand{\arraystretch}{1.1}
    \begin{tabular*}{\columnwidth}{@{\extracolsep{\fill}} lrrr}
        \toprule
        \textbf{Algorithm} & \textbf{SR (\%)} & \textbf{CR (\%)} & \textbf{Solve (ms)} \\
        \midrule
        DDP           & 100.0 & 100.0 & $25.53\pm1.92$ \\
        AL-DDP        & 100.0 & 100.0 & $48.17\pm3.44$ \\
        iLQR          & 100.0 & 100.0 & $20.73\pm0.51$ \\
        AL-iLQR       & 100.0 & 100.0 & $40.21\pm0.47$ \\
        SQP           & 100.0 & 100.0 & $16.59\pm0.03$ \\
        SQP (Geo.)    & 100.0 & 100.0 & $51.31\pm0.37$ \\
        \bottomrule
    \end{tabular*}
\end{table}

\subsection{Case Studies}

To complement the aggregate benchmark, we present two representative delivery missions that illustrate the framework's real-world utility in city districts where aerial mobility provides an advantage over ground transport.

\subsubsection{Package delivery from Orchard Road Store to Theater in Singapore (Case 1)}

\begin{figure}[tb]
\centering
\begin{subfigure}[b]{0.66\columnwidth}
    \centering
    \includegraphics[width=\textwidth]{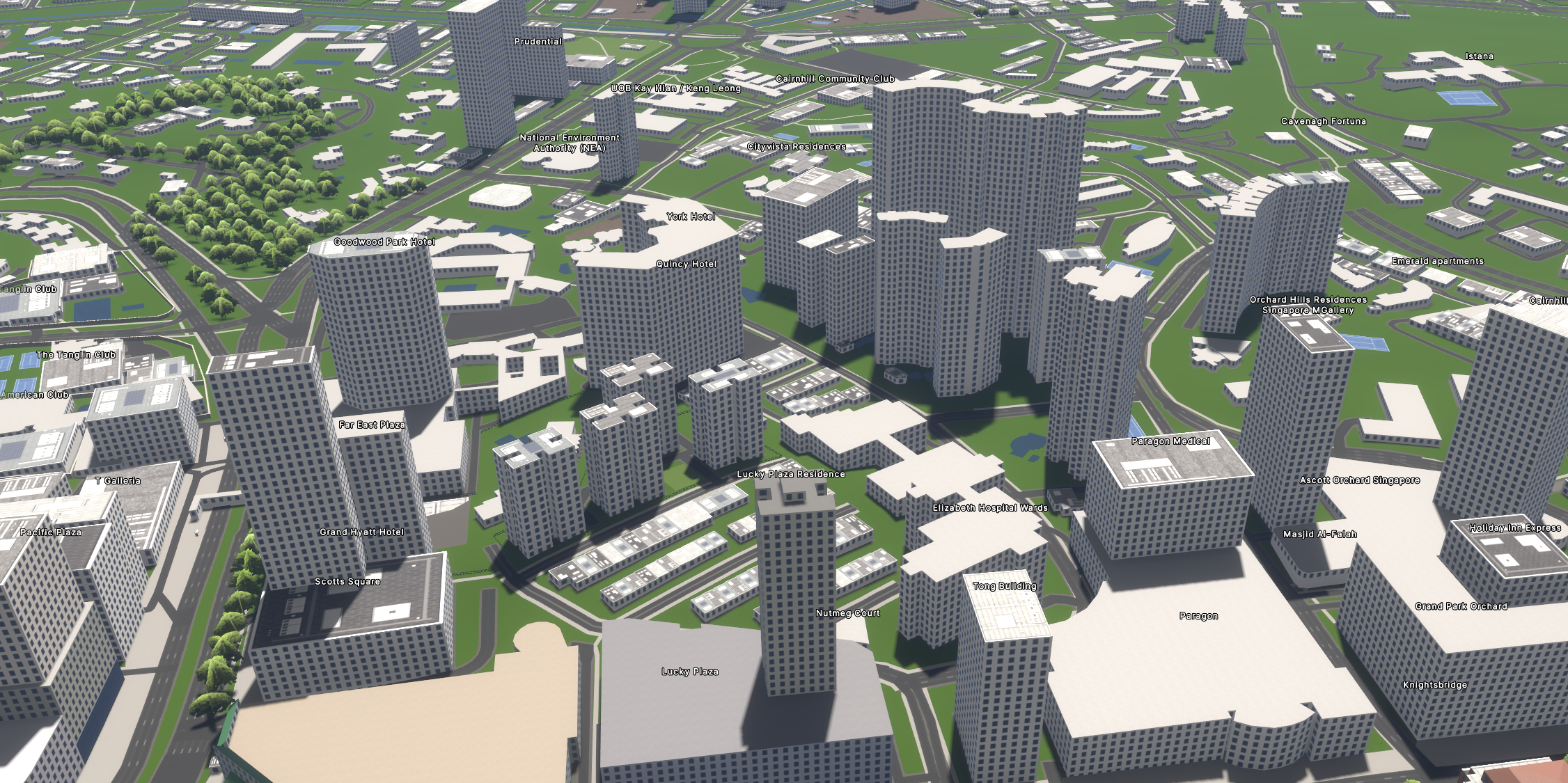}
    \caption{Orchard, Singapore}
\end{subfigure}%
\hfill
\begin{subfigure}[b]{0.33\columnwidth}
    \centering
    \includegraphics[width=\textwidth]{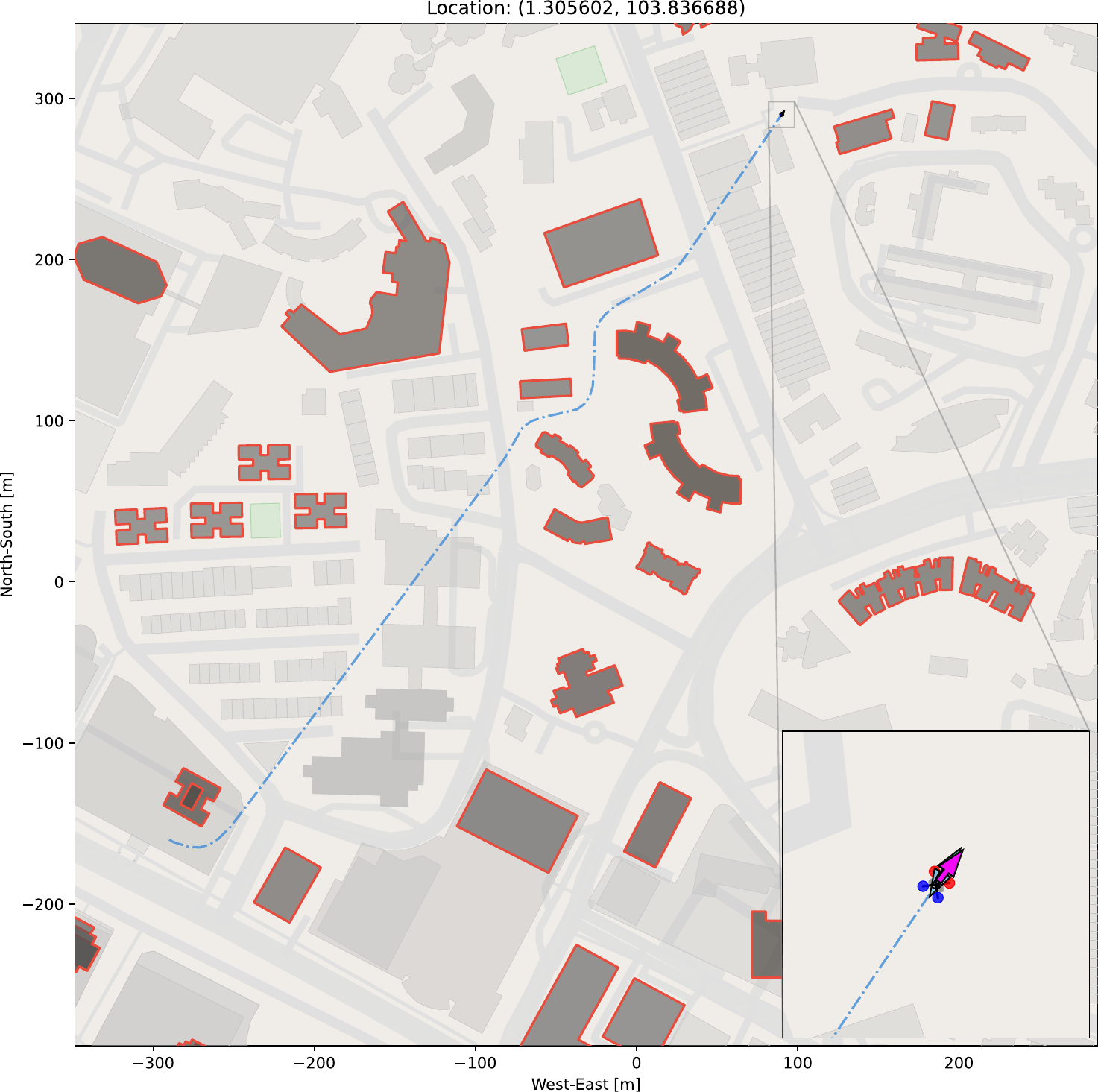}
    \caption{Environment}
\end{subfigure}

\vspace{4pt}

\begin{subfigure}[b]{0.33\columnwidth}
    \centering
    \includegraphics[width=\textwidth]{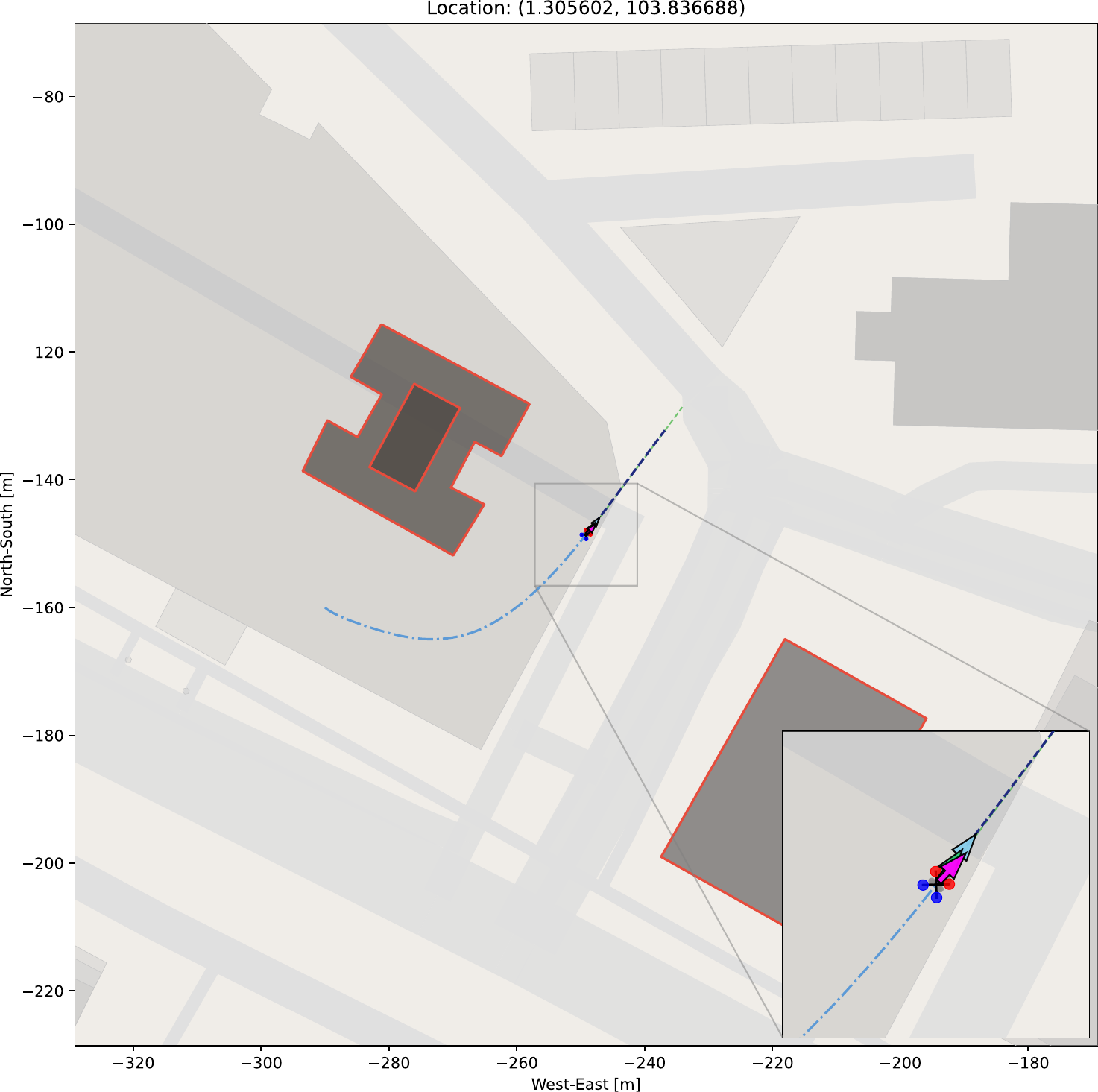}
    \caption{$t=5.7$\,s}
\end{subfigure}%
\hfill
\begin{subfigure}[b]{0.33\columnwidth}
    \centering
    \includegraphics[width=\textwidth]{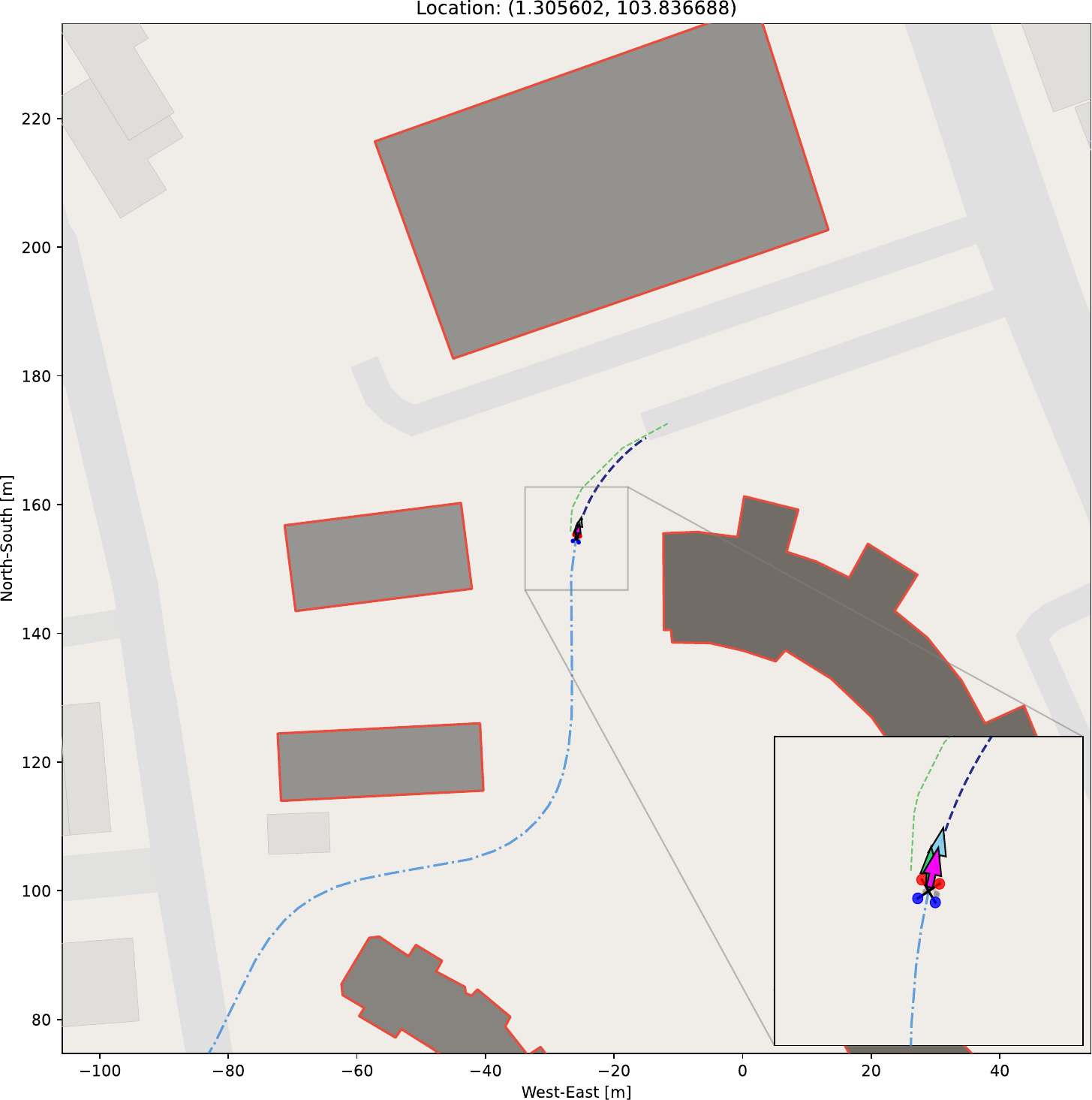}
    \caption{$t=43.2$\,s}
\end{subfigure}%
\hfill
\begin{subfigure}[b]{0.33\columnwidth}
    \centering
    \includegraphics[width=\textwidth]{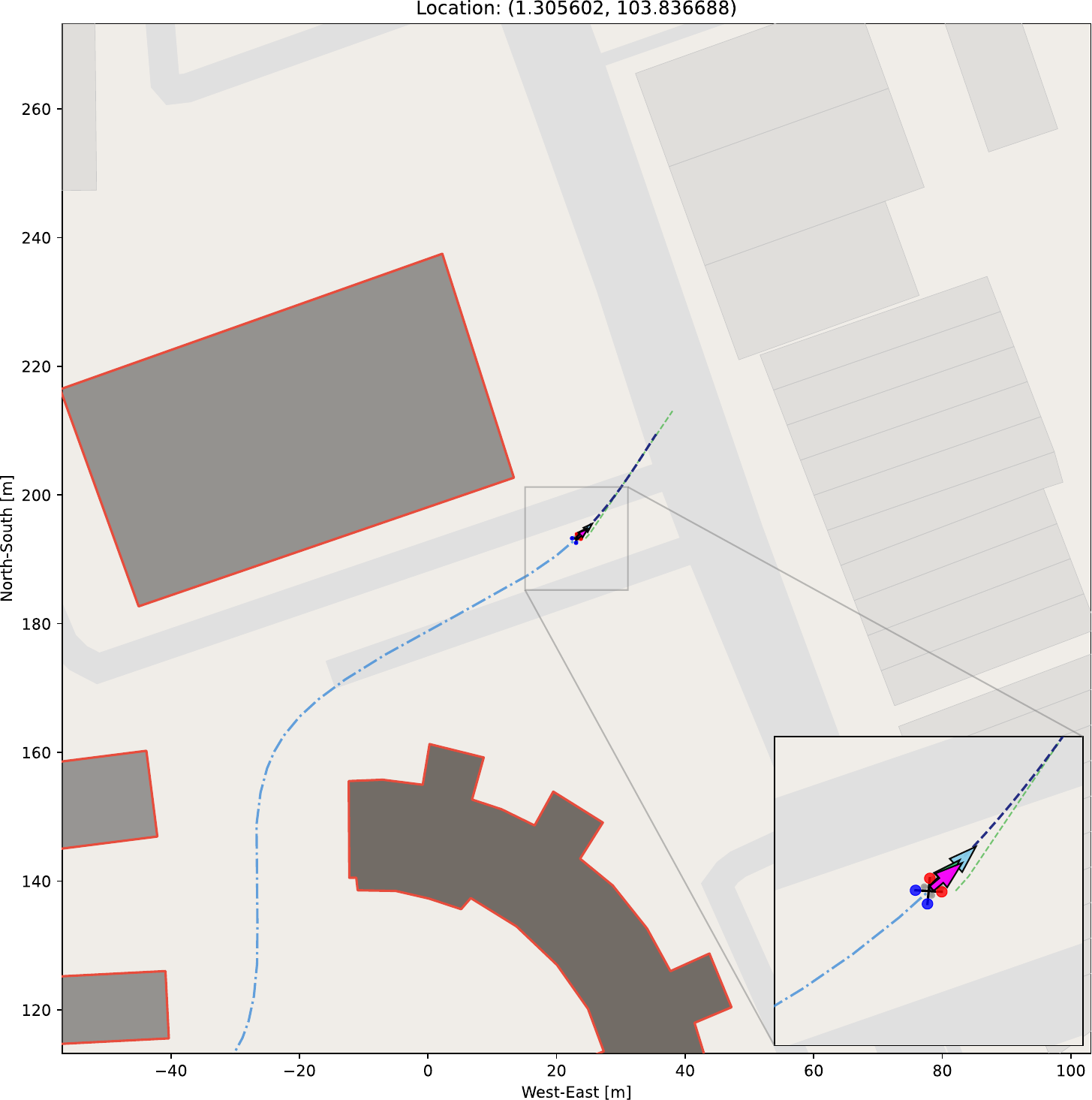}
    \caption{$t=49.9$\,s}
\end{subfigure}

\caption{4D trajectory flight mission for package delivery from Lucky Plaza departmental store to local theater.}\label{fig:orchard_trajectory}
\end{figure}

The mission covers approximately 1.3\,km of road through the Orchard Road shopping district, a journey of 4, 6, 5, and 18 minutes by motorcycle, car, bicycle, and foot, respectively, due to traffic signals and pedestrian crossings.
The LTP framework plans a direct aerial route at 50\,m altitude for approximately a minute of flight time, navigating between the dense high-rise facades along Orchard Road while respecting all geometric and dynamic constraints (Figure~\ref{fig:orchard_trajectory}).
The time saving would be most significant during peak shopping hours when ground-level congestion is highest.

\subsubsection{Food delivery from downtown Boston to Cambridge, Massachusetts (Case 2)}

\begin{figure}[tb]
\centering
\begin{subfigure}[b]{0.66\columnwidth}
    \centering
    \includegraphics[width=\textwidth]{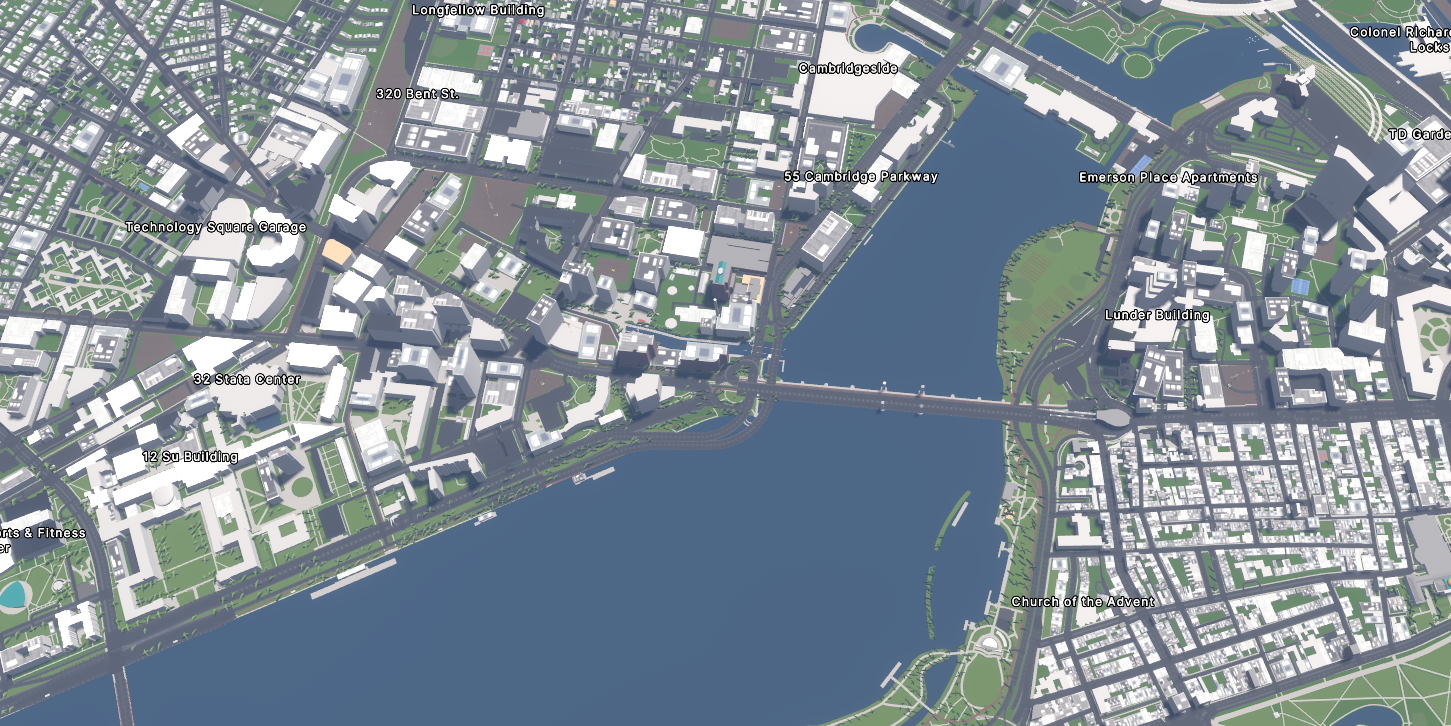}
    \caption{Charles River, Boston, Massachusetts}
\end{subfigure}%
\hfill
\begin{subfigure}[b]{0.33\columnwidth}
    \centering
    \includegraphics[width=\textwidth]{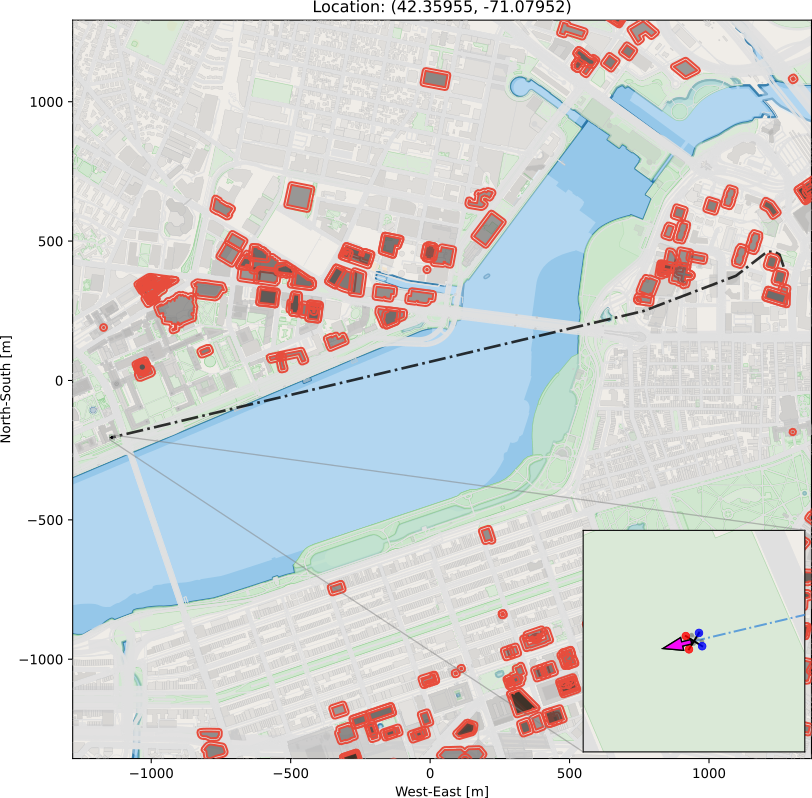}
    \caption{Environment}
\end{subfigure}

\vspace{4pt}

\begin{subfigure}[b]{0.33\columnwidth}
    \centering
    \includegraphics[width=\textwidth]{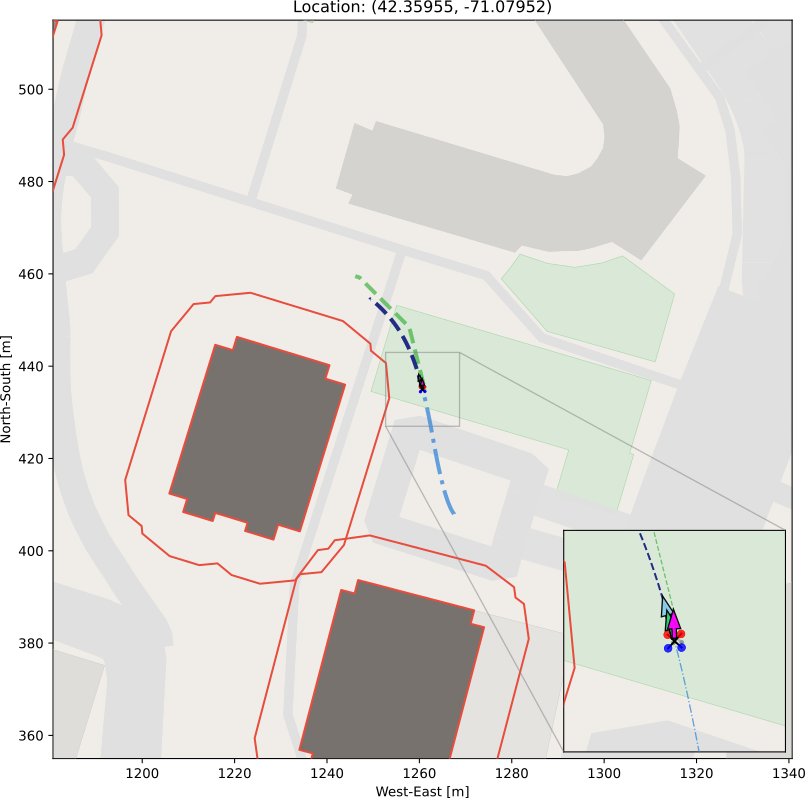}
    \caption{$t=3.6$\,s}
\end{subfigure}%
\hfill
\begin{subfigure}[b]{0.33\columnwidth}
    \centering
    \includegraphics[width=\textwidth]{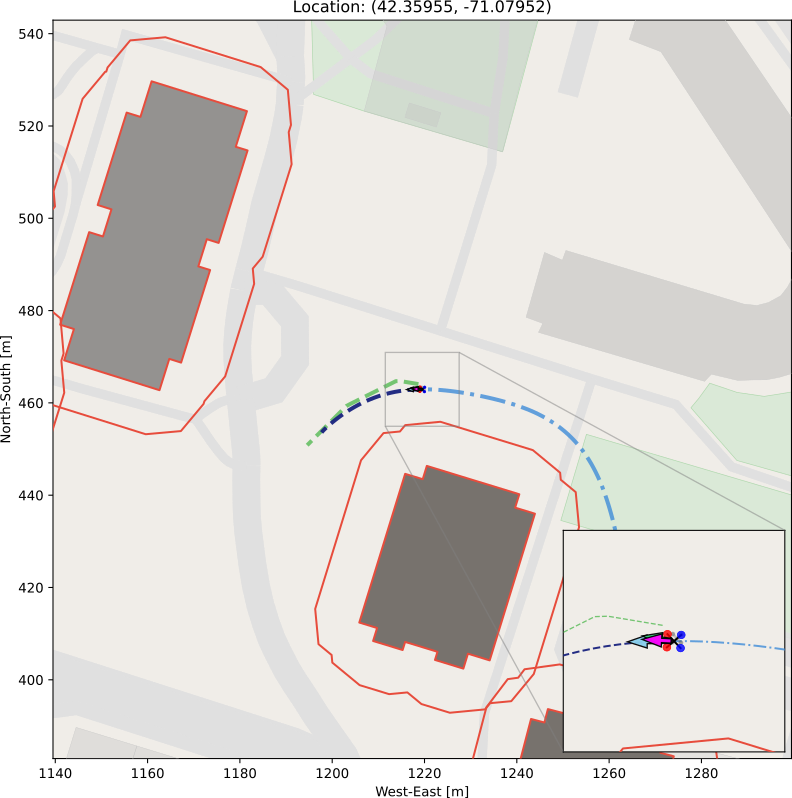}
    \caption{$t=8.9$\,s}
\end{subfigure}%
\hfill
\begin{subfigure}[b]{0.33\columnwidth}
    \centering
    \includegraphics[width=\textwidth]{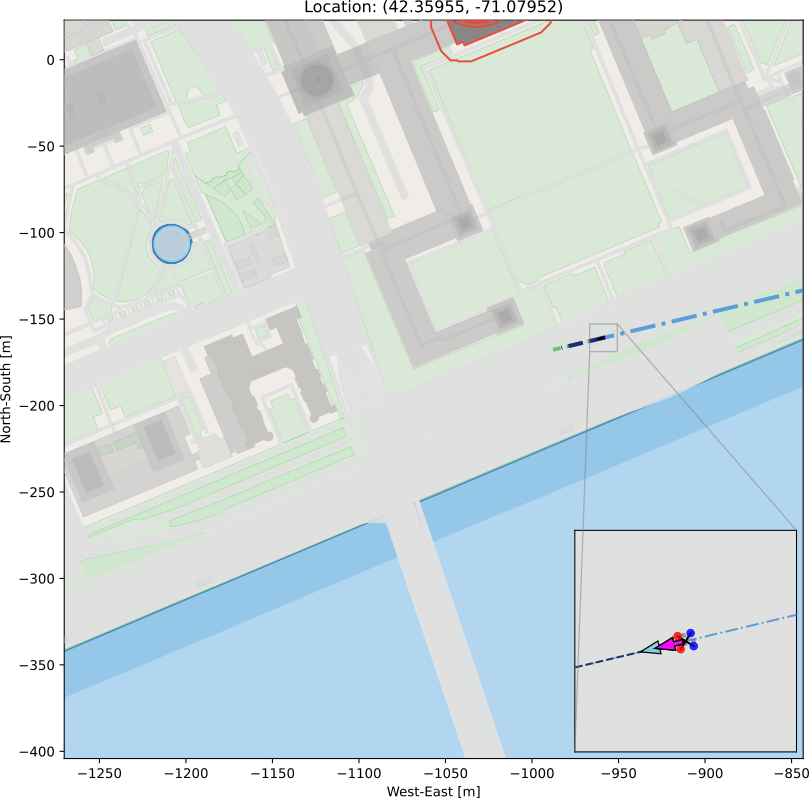}
    \caption{$t=221.7$\,s}
\end{subfigure}

\caption{4D trajectory flight mission for aerial food delivery from Mike's Sub to dinning hall.}\label{fig:boston_trajectory}
\end{figure}

The mission spans approximately 2.4\,km of road between downtown Boston and Cambridge, a route requiring 12, 15, 30, and 41 minutes by car, bicycle, public transit, and foot, respectively, as all ground routes must cross the Charles River via bridges.
The LTP framework plans an aerial trajectory that crosses directly over the Charles River at 50\,m altitude, bypassing bridge congestion entirely and completing the delivery in approximately 4 minutes of flight (Figure~\ref{fig:boston_trajectory}).
Because the majority of the flight path passes over open water rather than dense urban geometry, obstacle interactions are limited to the departure and arrival urban segments, yielding lower safety concerns, minimal geometric constraint activity, and heightened trajectory optimality.

\section{Conclusion}\label{sec:conclusion}

In this work, we presented the Last-Mile Trajectory Planning (LTP) framework, a real-time, dynamics-aware trajectory optimization architecture for small UAVs operating within an altitude-stratified urban air mobility channel. The framework couples a JAX-accelerated digital-twin model with an SQP optimizer that treats infrastructure and obstacle avoidance as a live separating-hyperplane constraint, regenerated at every inner solver call from the current candidate trajectory and the current digital environment. A variable-scale quadtree decomposition over the UTM channel further restricts each call to a locally relevant subset of geometry, so the per-call QP size remains compact and governed by local geometric density rather than by the size of the global map.

Across $1710$ flights spanning five real-world urban environments (Austin, Boston, Orchard Road, Marina Bay, and Hong Kong), $57$ routes, five target speeds up to $15$\,m/s, and different trajectory planning algorithms, the framework attained a $100\%$ success rate and a $100\%$ clearance rate, while solving each planning call in roughly $0.03$\,s on a CPU-only edge platform for a quadcopter model. Conventional DDP diverged on the highly nonlinear 12-state Newton-Euler quadcopter model, while iLQR was approximately $5.5\times$ slower than the base SQP planner and exhibited clearance violations at high speeds. Similarly, SQP variants without live geometric constraints lost clearance margin during aggressive turns. The geometric SQP variant introduced here addressed these limitations with no measurable increase in solve time, demonstrating the advantage of reasoning about obstacle avoidance and vehicle dynamics jointly within the SQP formulation. Two case studies (an Orchard Road delivery in Singapore and a cross-river delivery in Boston) further illustrated end-to-end utility in city districts where ground transport is fundamentally limited by traffic or topology.

Several directions for future work remain. First, integrating the planner into a comprehensive UTM stack, where strategic deconfliction, conformance monitoring, and contingency management are coordinated via digital-twins, would bridge the gap between airspace design and vehicle-level trajectory planning. Second, the linear-constraint formulation can naturally extend to virtual restrictions, such as dynamic geo-fences, noise-sensitive corridors, and environmental-impact zones, without modifying the core optimizer. Third, extending the constraint generator to handle moving obstacles (e.g., other UAM traffic or microclimate cells) will be critical for operations in fully shared airspace. Finally, formal safety verification of the closed-loop trajectory generation and tracking cascade remains a necessary step toward flight certification.

The source code is publicly available at \url{https://github.com/wzjoriv/Larp} along with the benchmark tested and case studies flights simulations.

\bibliographystyle{unsrt}
\bibliography{root}
\vspace{12pt}

\end{document}